\documentstyle[12pt,citesort]{article}
\newcommand{\sotimes}{\mathop{\otimes}_{s}}
\begin{document}
\begin{titlepage}
\renewcommand{\thefootnote}{\fnsymbol{footnote}}
\setcounter{footnote}{1}
\vspace*{50pt}

\begin{center}
{\Large \bf Integrable Boundary Conditions for the One-Dimensional
Hubbard Model} \\
\vspace{20pt}

{\Large
Masahiro {\sc Shiroishi}
\footnote{E-mail: siroisi@monet.phys.s.u-tokyo.ac.jp}
and Miki {\sc Wadati}%
\footnote{E-mail: wadati@monet.phys.s.u-tokyo.ac.jp}}

\vspace{20pt}

{\large\it Department of Physics, Graduate School of Science,\\
University of Tokyo, \\
Hongo 7--3--1, Bunkyo-ku, Tokyo 113, Japan}
\end{center}
\vspace{12pt}

\vfill

\begin{center}
{\bf Abstract} 
\end{center}

{\begin{quotation}
We discuss the integrable boundary conditions for the one-dimensional 
(1D) Hubbard Model in the framework of the Quantum Inverse Scattering 
Method (QISM). 
We use the fermionic ${R}$-matrix proposed by Olmedilla {\it et \ al.}
to treat the twisted periodic boundary condition and the open boundary
condition. We determine the most general form of the integrable
twisted periodic boundary condition by considering the symmetry matrix
of the fermionic ${R}$-matrix.  To find the integrable open boundary
condition, we shall solve the graded reflection equation, and find
there are two diagonal solutions, which correspond to a) the boundary 
chemical potential and b) the boundary magnetic field. Non-diagonal
solutions are obtained using the symmetry matrix of the fermionic 
${R}$-matrix and the covariance property of the graded reflection
equation. They can be interpreted as the ${SO(4)}$ rotations of 
the diagonal solutions.
\end{quotation}}

\vspace{50pt}

\end{titlepage}

\newpage

\begin{flushleft}
{\large \bf \S 1. Introduction}
\end{flushleft}
\setcounter{equation}{0}
\renewcommand{\theequation}{1.\arabic{equation}}
The one-dimensional (1D) Hubbard model is one of the most important
solvable models in condensed matter physics. The ground state energy
of the 1D Hubbard model
\begin{equation}
\hat{H}= - \sum_{m=1}^{N} \sum_{s = \uparrow \downarrow}
(c_{ms}^{\dagger} c_{m+1 s} + c_{m+1 s}^{\dagger} c_{m s}) 
+ U \sum_{m=1}^{N} ( n_{m \uparrow} - \frac{1}{2} ) 
( n_{m \downarrow} - \frac{1}{2} ), \label{eq.Hubbard}
\end{equation}
with periodic boundary condition
\begin{equation}
c^{\dagger}_{N+1s} = c_{1s}^{\dagger}, \ \ c_{N+1s} = c_{1s} \ \ 
\ \ (s = \uparrow \downarrow), \label{eq.PBC}
\end{equation}
was obtained by Lieb and Wu \cite{Lieb} by means of the coordinate 
Bethe ansatz method. Here ${c_{ms}^{\dagger}}$ and ${c_{ms}}$ are 
the fermionic creation and annihilation operators with spin 
${s (= \uparrow \downarrow)}$ at site ${m (= 1, 2, \cdots, N)}$, 
and ${n_{ms}}$ is the number density operator
\begin{equation}
n_{ms} = c_{ms}^{\dagger} c_{ms}. \label{eq.density}
\end{equation}
The parameter ${U}$ is the coupling constant 
describing the Coulomb interaction.
The bulk properties of the 1D Hubbard model have been clarified 
by analyzing the associated Bethe ansatz equations \cite{Korepin}.

As is well known, the Hamiltonian (\ref{eq.Hubbard}) enjoys 
two ${SU(2)}$ symmetries \cite{Heilmann,Yang1,Yang2,Pernici,Affleck}, 
which are the spin-${SU(2)}$ generated by
\begin{equation}
S^{+} = \sum_{m=1}^{N} c_{m \uparrow}^{\dagger} c_{m \downarrow}, \ \
S^{-} = \sum_{m=1}^{N} c_{m \downarrow}^{\dagger} c_{m \uparrow}, \ \ 
S^{z} = \frac{1}{2} \sum_{m=1}^{N} \left( n_{m \uparrow} - n_{m
\downarrow} \right), 
\label{eq.spinsu2}
\end{equation}
and the charge-${SU(2)}$ generated by
\begin{equation}
\eta^{+} = \sum_{m=1}^{N} (-1)^{m} c_{m \uparrow}^{\dagger} c_{m
\downarrow}^{\dagger}, \ \ 
\eta^{-} = \sum_{m=1}^{N} (-1)^{m} c_{m \downarrow} c_{m \uparrow}, \ \
\eta^{z} = \frac{1}{2} \sum_{m=1}^{N} \left( n_{m \uparrow} 
+ n_{m \downarrow} - 1 \right).
\end{equation}
For the consistency of the definition of the charge-${SU(2)}$, 
it is necessary to assume that the number of the lattice sites 
${N}$ is even. The spin-${SU(2)}$ and the charge-${SU(2)}$ are 
connected through the partial particle-hole transformation
\begin{equation}
c_{m \uparrow} \rightarrow c_{m \uparrow}, \ \ \ \ 
c_{m \downarrow} \rightarrow (-1)^{m} c_{m \downarrow}^{\dagger}, \ \ \ \ 
U \rightarrow -U. \label{eq.particlehole}
\end{equation} 

Since a constraint
\begin{equation}
S^{z} + \eta^{z} = {\rm integer}
\end{equation}
holds, the exact symmetry of the Hamiltonian (\ref{eq.Hubbard}) is 
${SO(4) = \left[ SU(2) \times SU(2) \right] / {\bf Z}_{2}}$. 
The ${SO(4)}$ symmetry may be the most fundamental property 
of the 1D Hubbard model that characterizes the various physical 
features of the model. For example, it was proved by E{\ss}ler 
{\it et al.} that the Bethe ansatz states of the 1D Hubbard model 
are incomplete and have to be  complemented 
by the ${SO(4)}$ symmetry \cite{Essler1,Essler2,Essler3}.  
E{\ss}ler and Korepin \cite{Essler4,Essler5} showed 
that the elementary excitations of the half-filled band constitute 
the multiplets of ${SO(4)}$.   

To discuss the bulk properties of the model, we usually assume 
the periodic boundary condition (\ref{eq.PBC}), since the bulk
physical quantities do not depend on the boundary condition. 
However, when we investigate the surface critical phenomena, 
or the transport properties, it is necessary to pay attention to 
the boundary condition.

The twisted periodic boundary condition for the Hubbard model was 
discussed by Shastry and Sutherland \cite{Shastry1,Martins}. 
They have shown that the coordinate Bethe ansatz method is applicable 
if we generalize the periodic boundary conditions to the twisted
periodic boundary condition
\begin{eqnarray}
& & c_{N+1 \uparrow}^{\dagger} = {\rm e}^{{\rm i} \phi} 
c_{1\uparrow}^{\dagger}, \ \ c_{N+1 \uparrow} = {\rm e}^{- {\rm i}
\phi} c_{1 \uparrow}, \nonumber \\
& & c_{N+1 \downarrow}^{\dagger} = {\rm e}^{{\rm i} \psi} 
c_{1\downarrow}^{\dagger}, \ \ c_{N+1 \downarrow} = {\rm e}^{ - {\rm
i} \psi} c_{1 \downarrow}, \label{eq.TPBC}
\end{eqnarray}
where the parameters ${\phi}$ and ${\psi}$ are twist angles. 
These changes of the boundary condition affect directly 
the transport properties such as 
the stiffness constants \cite{Shastry1,Martins}.

Recently the 1D Hubbard model with the open boundary condition has 
attracted much attention.  The corresponding Hamiltonian reads
\begin{eqnarray}
\hat{H}_{\rm open} & = & - \sum_{m=1}^{N-1} \sum_{s = \uparrow
\downarrow} (c_{m s}^{\dagger} c_{m+1 s} + c_{m+1 s}^{\dagger} c_{m s})
 + U \sum_{m=1}^{N} \left( n_{m \uparrow} - \frac{1}{2} \right)  
\left( n_{m \downarrow} - \frac{1}{2} \right)  \nonumber \\
& & -  p_{1 \uparrow} \left( n_{1 \uparrow} - \frac{1}{2} \right)  
- p_{1 \downarrow} \left( n_{1 \downarrow} - \frac{1}{2} \right)
 - p_{N \uparrow} \left( n_{N \uparrow} - \frac{1}{2} \right) - 
p_{N \downarrow} \left( n_{N \downarrow} - \frac{1}{2}  \right), 
\label{eq.BoundaryHubbard}
\end{eqnarray}
where ${p_{1s}, p_{Ns} \ (s = \uparrow \downarrow)}$ are the boundary
fields that specify the boundary conditions. 
In the case of vanishing boundary fields 
${p_{1s} = p_{Ns} =0 \ (s = \uparrow \downarrow)}$, 
this model was solved by Schulz \cite{Schulz} 
applying the coordinate Bethe ansatz method. 
More recently, Asakawa and Suzuki \cite{Asakawa1} extended the
solution to the boundary chemical potential case 
${p_{1 \uparrow} = p_{1 \downarrow}%
 = p_{N \uparrow} = p_{N \downarrow} = p}$. 
The present authors \cite{Shiroishi1} studied the consistency
conditions for the Bethe ansatz wave functions and classified the
possible forms of the boundary fields. 
It turned out that the boundary fields can be either of the form 
a) a boundary chemical potential: 
${p_{m \uparrow} = p_{m \downarrow} = p_{m},}$ 
or b) a boundary magnetic field: 
${p_{m \uparrow} = - p_{m \downarrow} = p_{m},}$
 where ${m = 1}$ or ${m=N}$. 

Since the boundary conditions at the different ends are independent, 
we have four integrable open boundary conditions \cite{Shiroishi1}: 
\begin{eqnarray}
& {\rm Case \ A}. & \ \ \ \ p_{1 \uparrow} = p_{1 \downarrow} = p_{1},
\ \ \ \ p_{N \uparrow} = p_{N \downarrow} = p_{N}, \nonumber \\
& {\rm Case \ B}. & \ \ \ \ p_{1 \uparrow} = - p_{1 \downarrow}=p_{1}, 
\ \ \ \ p_{N \uparrow} = - p_{N \downarrow} = p_{N}, \nonumber \\
& {\rm Case \ C}. & \ \ \ \ p_{1 \uparrow} = - p_{1 \downarrow} =p_{1}, 
\ \ p_{N \uparrow} = p_{N \downarrow} = p_{N}, \nonumber \\
& {\rm Case \ D}. & \ \ \ \ p_{1 \uparrow} = p_{1 \downarrow} = p_{1},
\ \  p_{N \uparrow} = - p_{N \downarrow} = p_{N}. \label{eq.4cases}
\end{eqnarray}
The corresponding Bethe ansatz equations were 
derived \cite{Shiroishi1,Deguchi}. 
  
The exact integrability of the 1D Hubbard model with periodic boundary
condition was established by Shastry \cite{Shastry2,Shastry3,Shastry4}
and Olmedilla {\it et al}. \cite{Olmedilla1,Olmedilla2}. 
The Jordan-Wigner transformation 
\begin{eqnarray}
& & c_{m \uparrow} = (\sigma_{1}^{z} \cdots \sigma_{m-1}^{z})
\sigma_{m}^{-}, 
\nonumber \\
& & c_{m \downarrow} = (\sigma_{1}^{z} \cdots \sigma_{N}^{z})
( \tau_{1}^{z} \cdots \tau_{m-1}^{z}) \tau_{m}^{-}, \label{eq.Jordan}
\end{eqnarray}
changes the fermionic Hamiltonian (\ref{eq.Hubbard}) into an
equivalent 
coupled spin model
\begin{equation}
H = \sum_{m=1}^{N} (\sigma_{m+1}^{+} \sigma_{m}^{-} + \sigma_{m}^{+}
\sigma_{m+1}^{-}) + \sum_{m=1}^{N} (\tau_{m+1}^{+} \tau_{m}^{-} 
+ \tau_{m}^{+} \tau_{m+1}^{-}) + \frac{U}{4}  \sum_{m=1}^{N} \sigma_{m}^{z}
\tau_{m}^{z}, \label{eq.Couple}
\end{equation}
where  $\sigma$ and $\tau$ are two species of the Pauli matrices 
commuting among each other, and
\begin{equation}
\sigma_{m}^{\pm} = \frac{1}{2} (\sigma_{m}^{x} \pm {\rm i}
\sigma_{m}^{y}), 
\ \ \ \ \tau_{m}^{\pm}   = \frac{1}{2} (\tau_{m}^{x} \pm {\rm i} 
\tau_{m}^{y}).
\end{equation}
Shastry constructed the ${L}$-operator and the ${R}$-matrix 
which satisfy the Yang-Baxter relation, ${``RLL = LLR",}$ 
for the equivalent coupled spin model (\ref{eq.Couple}). 
The present authors proved the Yang-Baxter equation, ${``RRR = RRR"}$,
for Shastry's ${R}$-matrix \cite{Shiroishi2,Shiroishi3,Shiroishi4}. 

It is well known that the Jordan-Wigner transformation
(\ref{eq.Jordan}) is not consistent with the periodic boundary
condition for the fermion operators (\ref{eq.PBC}). 
In fact the periodic boundary condition for the fermionic operators 
(\ref{eq.PBC}) corresponds to a sector dependent twisted boundary
condition for the coupled spin model (\ref{eq.Couple})
\begin{equation}
\left( \prod_{m=1}^{N} \sigma_{m}^{z} \right) \sigma_{N+1}^{\pm} 
= \sigma_{1}^{\pm}, \ \ \ \ 
\left( \prod_{m=1}^{N} \tau_{m}^{z} \right) \tau_{N+1}^{\pm} 
= \tau_{1}^{\pm}.
\end{equation}
Hence it is more natural to use the fermionic formulation 
developed by Olmedilla {\it et al.} \cite{Olmedilla1,Olmedilla2}, 
when we discuss the boundary conditions for the 1D Hubbard model. 
They proposed the fermionic ${L}$-operator and the fermionic
${R}$-matrix which satisfy the graded Yang-Baxter relation. 
The transfer matrix which corresponds to the fermionic 1D Hubbard
model with the periodic boundary condition is constructed by taking 
the supertrace of the monodromy matrix \cite{Olmedilla1,Olmedilla2}. 

In this paper, as a first problem, we shall generalize their results 
to the case of the twisted periodic boundary condition. 
We shall find the symmetry matrix of the fermionic ${R}$-matrix, 
which is the constant solution of the graded Yang-Baxter relation.  
It is closely related to the ${SO(4)}$ symmetry of the Hamiltonian 
(\ref{eq.Hubbard}). The symmetry matrix can be inserted into the
transfer matrix without breaking the integrability \cite{Sklyanin}.  
We shall show that the periodic boundary condition is twisted 
by the insertion of the symmetry matrix. 
 
A general method to prove the exact integrability of the model 
with open boundaries was developed by Sklyanin \cite{Sklyanin}.  
Through the Jordan-Wigner transformation, 
the Hamiltonian (\ref{eq.BoundaryHubbard}) is transformed into 
\begin{eqnarray}
H_{\rm open} & = & \sum_{m=1}^{N-1} \left( \sigma_{m}^{+}
\sigma_{m+1}^{-} + \sigma_{m}^{-} \sigma_{m+1}^{+} \right) 
+ \sum_{m=1}^{N-1} \left( \tau_{m}^{+} \tau_{m+1}^{-} + 
\tau_{m}^{-} \tau_{m+1}^{+} \right) + \frac{U}{4} \sum_{m=1}^{N} 
\sigma_{m}^{z} \tau_{m}^{z} \nonumber \\
& & - \frac{p_{1 \uparrow}}{2} \sigma_{1}^{z} - \frac{p_{N \uparrow}}{2} 
\sigma_{N}^{z} - \frac{p_{1 \downarrow}}{2} 
\tau_{1}^{z} - \frac{p_{N \downarrow}}{2} \tau_{N}^{z}. 
\label{eq.BoundaryCoupledSpin}
\end{eqnarray}
Zhou \cite{Zhou} applied the Sklyanin's formalism to investigate 
the exact integrability of the model (\ref{eq.BoundaryCoupledSpin}). 
He formulated the reflection equation in terms of Shastry's
${R}$-matrix and found a solution which corresponds to the boundary 
chemical potential. Subsequently, the solution corresponding to the 
boundary magnetic field was also found \cite{Guan}. 

In this paper, as a second problem, we shall formulate the graded
reflection equation in terms of the fermionic ${R}$-matrix. 
It is shown explicitly that there are only two diagonal solutions 
for the graded reflection equation, which correspond to a) the
boundary chemical potentials and b) the boundary magnetic fields. 
In this way, the integrability of the 1D Hubbard model with boundary 
fields (\ref{eq.BoundaryHubbard}) is proved. 
Moreover, making use of the covariance property of the reflection
equation, we construct two non-diagonal solutions. 
It is shown that the partial particle-hole transformation for 
the fermionic ${R}$-matrix relates these solutions.      

\vspace{20pt}
\begin{flushleft}
{\large \bf \S 2. Fermionic ${\mbox{\boldmath $R$}}$-Matrix and %
Graded Yang-Baxter Relation }
\end{flushleft}
\setcounter{equation}{0}
\renewcommand{\theequation}{2.\arabic{equation}}
Let us recall the fermionic formulation of the integrability 
of the 1D Hubbard model \cite{Olmedilla1,Olmedilla2}. 
The fermionic ${L}$-operator is given by
\begin{equation}
{\cal{L}}_{m}(\theta) =  {\cal{L}}(\theta) ( {\cal{L}}_{m
\uparrow}(\theta) 
\sotimes {\cal{L}}_{m \downarrow}(\theta) ) 
{\cal{L}}(\theta), \label{eq.fermionicL}
\end{equation}
where
\begin{equation}
{\cal{L}} (\theta) = \cosh (\frac{h}{2}) I + 
\sinh (\frac{h}{2}) \sigma^{z} \otimes \sigma^{z},
\end{equation}
and
\begin{equation}
{\cal{L}}_{m \uparrow}(\theta) 
= \left(
         \begin{array}{cc}
       - f_{m \uparrow}(\theta) &  {\rm i} c_{m \uparrow} \\
         c_{m \uparrow}^{\dagger} & g_{m \uparrow}(\theta) \\
          \end{array}
   \right), \ \ \ \  
{\cal{L}}_{m \downarrow}(\theta) 
=  \left(
     \begin{array}{cc}
      f_{m \downarrow}(\theta) & c_{m \downarrow} \\
    {\rm i} c_{m \downarrow}^{\dagger} & - g_{m \downarrow}(\theta)  
    \end{array}
    \right).
\end{equation}
Here ${I}$ is the 4 ${\times}$ 4 identity matrix, 
${\otimes}$ means the usual direct product, and
\begin{eqnarray}
f_{m s}(\theta) & = & \sin \theta - \left\{ \sin \theta - {\rm i} 
\cos \theta \right\} n_{m s}, \nonumber \\
g_{m s}(\theta) & = & \cos \theta - \left\{ \cos \theta + {\rm i} 
\sin \theta \right\} n_{m s}.
\end{eqnarray}
The symbol ${\displaystyle \sotimes}$ denotes the Grassmann (graded) 
direct product
\begin{eqnarray}
{[ A \sotimes B ]}_{\alpha \gamma, \beta \delta} 
&=& (-1)^{\left[ P(\alpha) + P(\beta) \right] P(\gamma)} 
A_{\alpha \beta} B_{\gamma \delta}, \nonumber \\
P(1) &=& 0, \ \ P(2) = 1. \label{eq.Grassmann}
\end{eqnarray}
It is sometimes convenient to write the fermionic ${L}$-operator 
explicitly, 
\begin{equation}
{\cal{L}}_{m}(\theta) = 
\left(
 \begin{array}{cccc}
- {\rm e}^{h} f_{m \uparrow}(\theta) f_{m \downarrow}(\theta) &  
 - f_{m \uparrow}(\theta) c_{m \downarrow} &
 {\rm i} c_{m \uparrow} f_{m \downarrow}(\theta) &
 {\rm i} {\rm e}^{h} c_{m \uparrow} c_{m \downarrow} \\
 - {\rm i} f_{m \uparrow}(\theta) c_{m \downarrow}^{\dagger} &
  {\rm e}^{-h} f_{m \uparrow}(\theta) g_{m \downarrow}(\theta) &
  {\rm e}^{-h} c_{m \uparrow} c_{m \downarrow}^{\dagger} &
 {\rm i} c_{m \uparrow} g_{m \downarrow}(\theta) \\
  c_{m \uparrow}^{\dagger} f_{m \downarrow}(\theta) &
  {\rm e}^{-h} c_{m \uparrow}^{\dagger} c_{m \downarrow} &
  {\rm e}^{-h} g_{m \uparrow}(\theta) f_{m \downarrow}(\theta) &
  g_{m \uparrow}(\theta) c_{m \downarrow} \\
  - {\rm i} {\rm e}^{h} c_{m \uparrow}^{\dagger} 
c_{m \downarrow}^{\dagger} &
  c_{m \uparrow}^{\dagger} g_{m \downarrow}(\theta) &
  {\rm i} g_{m \uparrow}(\theta) c_{m \downarrow}^{\dagger} &
  - {\rm e}^{h} g_{m \uparrow}(\theta) g_{m \downarrow}(\theta)
 \end{array}
\right).
\end{equation}
The parameter ${h}$ should be considered as a function of 
the spectral parameter ${\theta}$ and the Coulomb coupling 
constant ${U}$ through the relation
\begin{equation}
\frac{\sinh 2h}{\sin 2 \theta} = \frac{U}{4}. 
\label{eq.constraint1}
\end{equation}
The fermionic ${L}$-operator fulfills the graded Yang-Baxter 
relation \cite{Olmedilla1} 
\begin{equation}
\check{\cal{R}}_{12}(\theta_1,\theta_2) 
[ {\cal{L}}_{m} (\theta_1) \sotimes {\cal{L}}_{m} (\theta_2) ] 
= [ {\cal{L}}_{m} (\theta_2) \sotimes {\cal{L}}_{m} (\theta_1) ] 
{\check{\cal{R}}}_{12} (\theta_1,\theta_2). \label{eq.GYBR}
\end{equation}
Here the parity of the Grassmann direct product 
${\displaystyle \sotimes}$ is assigned as
\begin{equation}
P(1) = P(4) = 0, \ \ P(2) = P(3) = 1. \label{eq.parity}
\end{equation}
In the graded Yang-Baxter relation (\ref{eq.GYBR}), the constraints 
\begin{equation}
\frac{\sinh 2 h_{1}}{\sin 2 \theta_{1}} = 
\frac{\sinh 2 h_{2}}{\sin 2 \theta_{2}} = \frac{U}{4}, 
\label{eq.constraint2}
\end{equation}
are assumed.
The matrix elements of the fermionic ${R}$-matrix 
${{\check{\cal{R}}}_{12}(\theta_1,\theta_2)}$ are given in ref. 22. 
For later use, we introduce an equivalent fermionic ${R}$-matrix
\begin{equation}
{\cal{R}}_{12}(\theta_1,\theta_2) \equiv {\cal{P}}_{12} 
\check{\cal{R}}_{12}(\theta_1,\theta_2),
\end{equation}
where ${{\cal{P}}_{12}}$ is the graded permutation
\begin{equation}
{\cal{P}}_{\alpha \gamma, \beta \delta} = 
(-1)^{P(\alpha) P(\gamma)} \delta_{\alpha \delta} \delta_{\gamma \beta}.
\end{equation}
Using the fermionic ${R}$-matrix ${{\cal R}_{12}(\theta_1,\theta_2)}$,
the graded Yang-Baxter relation (\ref{eq.GYBR}) can be expressed as
\begin{equation}
 {\cal R}_{12}(\theta_1,\theta_2) 
\left( {\cal L}_{m}(\theta_{1}) \sotimes I \right) 
\left( I \sotimes {\cal L}_{m}(\theta_{2}) \right) 
= \left( I \sotimes {\cal L}_{m}(\theta_{2}) \right) 
\left( {\cal L}_{m}(\theta_{1}) \sotimes I \right) 
{\cal R}_{12}(\theta_1,\theta_2).
\end{equation}
We parametrize the matrix elements of the fermionic ${R}$-matrix 
as follows
{\scriptsize
\begin{eqnarray}
& & {\cal R}_{12}(\theta_1,\theta_2) = \nonumber \\
& & \left( 
    \begin{array}{cccccccccccccccc}
 a^{+}&0&0&0&0&0&0&0&0&0&0&0&0&0&0&0 \\
 0&-{\rm i}b^{+}&0&0&e&0&0&0&0&0&0&0&0&0&0&0 \\
 0&0&-{\rm i}b^{+}&0&0&0&0&0&e&0&0&0&0&0&0&0 \\
 0&0&0&-c^{+}&0&0&{\rm i}f&0&0&-{\rm i}f&0&0&d^{+}&0&0&0 \\
 0&e&0&0&{\rm i}b^{-}&0&0&0&0&0&0&0&0&0&0&0 \\
 0&0&0&0&0&-a^{-}&0&0&0&0&0&0&0&0&0&0 \\
 0&0&0&{\rm i}f&0&0&c^{-}&0&0&-d^{-}&0&0&-{\rm i}f&0&0&0 \\
 0&0&0&0&0&0&0&{\rm i}b^{-}&0&0&0&0&0&e&0&0 \\
 0&0&e&0&0&0&0&0&{\rm i}b^{-}&0&0&0&0&0&0&0 \\
 0&0&0&-{\rm i}f&0&0&-d^{-}&0&0&c^{-}&0&0&{\rm i}f&0&0&0 \\
 0&0&0&0&0&0&0&0&0&0&-a^{-}&0&0&0&0&0 \\
 0&0&0&0&0&0&0&0&0&0&0&{\rm i}b^{-}&0&0&e&0 \\
 0&0&0&d^{+}&0&0&-{\rm i}f&0&0&{\rm i}f&0&0&-c^{+}&0&0&0 \\
 0&0&0&0&0&0&0&e&0&0&0&0&0&-{\rm i}b^{+}&0&0 \\
 0&0&0&0&0&0&0&0&0&0&0&e&0&0&-{\rm i}b^{+}&0 \\
 0&0&0&0&0&0&0&0&0&0&0&0&0&0&0&a^{+} \\
      \end{array}
      \right), \nonumber \\ \label{eq.Rmatrix}
\end{eqnarray}
}
where
\begin{eqnarray}
a^{\pm} & = & \cos^2 (\theta_1 - \theta_2) \left\{ 1 \pm \tanh (h_1 - h_2) 
\frac{ \cos (\theta_1 + \theta_2)}{\cos (\theta_1 - \theta_2)} \right\}, 
\nonumber \\
b^{\pm} & = & \sin (\theta_1 - \theta_2) \cos (\theta_1 - \theta_2) 
\left\{  1 \pm \tanh (h_1 - h_2) \frac{\sin (\theta_1 + \theta_2)}
{\sin (\theta_1 - \theta_2)} \right\} \nonumber \\
        & = & \sin (\theta_1 - \theta_2) \cos (\theta_1 - \theta_2) 
\left\{ 1 \pm \tanh (h_1 + h_2) \frac{ \cos (\theta_1 + \theta_2)}
{\cos (\theta_1 - \theta_2)} \right\}, \nonumber \\
c^{\pm} & = & \sin^2 (\theta_1 - \theta_2) \left\{ 1 \pm \tanh (h_1 + h_2) 
\frac{\sin (\theta_1 + \theta_2)}{\sin (\theta_1 - \theta_2)} \right\}, 
\nonumber \\
d^{\pm} & = & 1 \pm \tanh (h_1 - h_2) 
\frac{\cos (\theta_1 - \theta_2)}{\cos (\theta_1 + \theta_2)}, 
\nonumber \\
        & = & 1 \pm \tanh(h_1 + h_2) 
\frac{\sin (\theta_1 - \theta_2)}{\sin (\theta_1 + \theta_2)}, 
\nonumber \\  
e       & = & \frac{\cos (\theta_1 - \theta_2)}{\cosh (h_1 - h_2)}, \ \ \ \ 
f =  \frac{\sin (\theta_1 - \theta_2)}{\cosh (h_1 + h_2)}.
\end{eqnarray} 
The second equalities for the Boltzmann weights ${b^{\pm}}$ and
${d^{\pm}}$ are due to the constraints (\ref{eq.constraint2}), 
or equivalently the relation 
\begin{equation}
\frac{\tanh (h_1 - h_2)}{\tanh (h_1 + h_2)} 
= \frac{\tan (\theta_1 - \theta_2)}{\tan (\theta_1 + \theta_2)}.
\end{equation} 
We note some useful relations among the Boltzmann weights 
\cite{Olmedilla1},
\begin{eqnarray}
& & d^{\pm} = a^{\pm} + c^{\pm}, \ \ 
d^{+} d^{-} = e^{2} + f^{2}, \nonumber \\
& & e^{2} = a^{+} a^{-} + b^{+} b^{-}, \ \ 
f^{2} = b^{+} b^{-} + c^{+} c^{-}.
\end{eqnarray}
The monodromy matrix is defined as the ordered product of 
the fermionic ${L}$-operators
\begin{equation}
T(\theta) = \prod_{m=1}^{N \atop \longleftarrow} {\cal L}_{m}(\theta) 
= {\cal L}_{N}(\theta) \cdots {\cal L}_{1}(\theta). \label{eq.monodromy}
\end{equation} 
 From the (local) graded Yang-Baxter relation (\ref{eq.GYBR}), 
we have the global relation for the monodromy matrix
\begin{eqnarray}
& & {\check{\cal R}}_{12}(\theta_1,\theta_2) 
[ T(\theta_1) \sotimes T(\theta_2) ] = [ T(\theta_2) \sotimes T(\theta_1) ] 
{\check{\cal R}}_{12}(\theta_1,\theta_2), \label{eq.globalGYBR}
\end{eqnarray}
or equivalently
\begin{eqnarray}
& & {\cal R}_{12}(\theta_1,\theta_2) \stackrel{1}{T}(\theta_1) 
\stackrel{2}{T}(\theta_2) = \stackrel{2}{T}(\theta_2) 
\stackrel{1}{T}(\theta_1) {\cal R}_{12}(\theta_1,\theta_2), 
\label{eq.globalGYBR2}
\end{eqnarray}
where
\begin{equation}
\stackrel{1}{T}(\theta)  \equiv {T}(\theta_1) \sotimes I, \ \ 
\stackrel{2}{T}(\theta)  \equiv  I \sotimes {T}(\theta_2).
\end{equation}
In other words, the monodromy matrix (\ref{eq.monodromy}) is 
a representation of the associative algebra ${T}$ defined by 
(\ref{eq.globalGYBR}) or (\ref{eq.globalGYBR2}).

By taking the supertrace of (\ref{eq.globalGYBR}) or
(\ref{eq.globalGYBR2}), 
we find that the transfer matrix 
\begin{equation}
t(\theta) = {\rm str} T(\theta) \equiv {\rm tr} 
\left\{ \left( \sigma^{z} \otimes \sigma^{z} \right) 
T(\theta) \right\} \label{eq.supertrace}
\end{equation}
constitutes a commuting family
\begin{equation}
\left[ t(\theta_{1}), t(\theta_{2}) \right] = 0.
\end{equation}
Olmedilla {\it et al.} \cite{Olmedilla1} showed that 
the Hamiltonian (\ref{eq.Hubbard}) under the periodic 
boundary condition (\ref{eq.PBC}) can be obtained by the 
series expansion of the transfer matrix ${t(\theta)}$ around ${\theta=0}$.

Now we list some properties enjoyed by 
the fermionic ${R}$-matrix ${{\cal R}_{12}(\theta_1,\theta_2)}$. \\
(1) Regularity (Initial condition): 
\begin{equation}
  {\cal R}_{12}(\theta_{0},\theta_{0}) = {\cal P}_{12}.
\end{equation}
(2) Graded Yang-Baxter equation:
\begin{equation}
{\cal R}_{12}(\theta_1,\theta_2) {\cal R}_{13}(\theta_1,\theta_3) 
{\cal R}_{23}(\theta_2,\theta_3) 
= {\cal R}_{23}(\theta_2,\theta_3) {\cal R}_{13}(\theta_1,\theta_3) 
{\cal R}_{12}(\theta_1,\theta_2),
\end{equation}
where we assume the constraints
\begin{equation}
\frac{\sinh 2 h_1}{\sin 2 \theta_1} = \frac{\sinh 2 h_2}{\sin 2
\theta_2} 
= \frac{\sinh 2 h_3}{\sin 2 \theta_3} = \frac{U}{4}.
\end{equation}
(3) Unitarity: 
\begin{equation}
{\cal R}_{12}(\theta_1,\theta_2) {\cal R}_{21}(\theta_2,\theta_1) 
= \rho (\theta_1,\theta_2) \ I , \label{eq.unitarity}
\end{equation}
where 
\begin{equation}
{\cal R}_{21}(\theta_2,\theta_1) \equiv {\cal P}_{12} 
{\cal R}_{12}(\theta_2,\theta_1) {\cal P}_{12},
\end{equation} 
 and
\begin{equation}
\rho(\theta_1,\theta_2) =  \cos^{2}(\theta_1 - \theta_2) 
\big\{ \cos^{2}(\theta_1 - \theta_2) 
- \tanh^2(h_1 - h_2) \cos^2(\theta_1 + \theta_2) \big\}.
\end{equation}
(4) Crossing unitarity: 
\begin{equation}
{\cal R}_{12}^{{\rm st}_1}(-\theta_1, \theta_2) 
{\cal R}_{21}^{{\rm st}_1}(\theta_2,-\theta_1)  
= {\cal R}_{12}^{{\rm st}_2}(-\theta_1,\theta_2) 
{\cal R}_{21}^{{\rm st}_2}(\theta_2,-\theta_1)  
= \tilde{\rho}(\theta_1,\theta_2) \ I, 
\label{eq.crossingunitarity}
\end{equation}
where
\begin{equation}
\tilde{\rho} (\theta_1,\theta_2) = \sin^{2} (\theta_1 + \theta_2) 
\big\{ \sin^{2}(\theta_1 + \theta_2) 
- \tanh^{2} (h_1 - h_2) \sin^{2} (\theta_1 - \theta_2) \big\}.
\end{equation}
Here the supertransposition acts on the ${4 \times 4}$ matrix 
with the parity (\ref{eq.parity}) as follows
\begin{equation}
{\left( \begin{array}{cccc}
           D_{11} & C_{11} & C_{12} & D_{12} \\
           B_{11} & A_{11} & A_{12} & B_{12} \\
           B_{21} & A_{21} & A_{22} & B_{22} \\
           D_{21} & C_{21} & C_{22} & D_{22} 
       \end{array}
\right)}^{\rm st} = \left( \begin{array}{cccc}
           D_{11} & -B_{11} & -B_{21} & D_{21} \\
           C_{11} &  A_{11} &  A_{21} & C_{21} \\
           C_{12} &  A_{12} &  A_{22} & C_{22} \\
           D_{12} & -B_{12} & -B_{22} & D_{22} 
       \end{array}
\right)
\end{equation}
In (\ref{eq.crossingunitarity}), ${{\rm st}_{j}}$ means 
the supertransposition with respect to the ${j}$-th space. 

We further note some properties of the ${R}$-matrix,
\begin{eqnarray}
{\cal R}_{12}(\theta_1,\theta_2) 
& = & {\cal R}_{12}(\theta_1 + \frac{\pi}{2}, \theta_2 + \frac{\pi}{2}) 
\nonumber \\
& = & {\cal R}_{12}(\theta_1 - \frac{\pi}{2},\theta_2 - \frac{\pi}{2}), 
\label{eq.pi2} \\
{\cal R}_{12}(\theta_1,\theta_2)^{{\rm st}_{1}, \overline{\rm st}_{2}}
& = & {\cal R}_{12}(\theta_1,\theta_2)^{\overline{\rm st}_{1},
{\rm st}_{2}}. \label{eq.Tsymmetry}
\end{eqnarray}
Here ${\overline{\rm st}_{j}}$ stands for the inverse 
of the supertransposition ${{\rm st}_{j}}$,
\begin{equation}
\left( X^{\rm st} \right)^{\overline{\rm st}} = 
\left( X^{\overline{\rm st}} \right)^{\rm st} = X.
\end{equation}
 We remark that ${{\cal R}_{12}(\theta_1,\theta_2)^{{\rm st}_{1}, %
\overline{\rm st}_{2}}}$ can be obtained from the ${R}$-matrix 
${{\cal R}_{12}(\theta_1,\theta_2)}$ by exchanging the Boltzmann
weights as ${{\rm i} f \leftrightarrow - {\rm i} f}$. 

The graded tensor product is assumed in the above relations. 
The graded Yang-Baxter equation, for instance, is expressed as 
follows \cite{Kulish}
\begin{eqnarray}
& & {\cal R}_{ab;a'b'}(\theta_1,\theta_2) 
{\cal R}_{a'c;jc'}(\theta_1,\theta_3) 
{\cal R}_{b'c';kl}(\theta_2,\theta_3) (-)^{P(b')(P(j) + P(a'))} 
\nonumber \\
& & = {\cal R}_{bc;b'c'}(\theta_2,\theta_3) 
{\cal R}_{ac';a'l}(\theta_1,\theta_3) 
{\cal R}_{a'b';jk}(\theta_1,\theta_2) (-)^{P(b')(P(a) + P(a'))}.
\end{eqnarray} 
The authors gave a proof of the Yang-Baxter equation 
for Shastry's ${R}$-matrix \cite{Shiroishi2,Shiroishi3,Shiroishi4}. 
 We can show that the graded Yang-Baxter equation for the 
fermionic ${R}$-matrix is equivalent to the Yang-Baxter equation 
for Shastry's ${R}$-matrix.  

\vspace{30pt}
\begin{flushleft}
{\large \bf \S 3. Symmetry of the Fermionic ${\mbox{\boldmath %
$R$}}$-Matrix and Twisted Periodic Boundary Condition}
\end{flushleft}
\setcounter{equation}{0}
\renewcommand{\theequation}{3.\arabic{equation}}
In this section  we shall discuss two important symmetries of the 
fermionic ${R}$-matrix.
 The symmetry of the fermionic ${R}$-matrix is defined by the constant
matrix ${M = (M_{ij})}$ that satisfies
\begin{equation}
\left[ \check{\cal R}_{12}(\theta_1,\theta_2), M \sotimes M \right] 
= 0, \label{eq.symmetrydef}
\end{equation}
or equivalently,
\begin{equation}
{\cal R}_{12}(\theta_1,\theta_2) ( M \sotimes I ) ( I \sotimes M ) 
= ( I \sotimes M ) ( M \sotimes I ) {\cal R}_{12}(\theta_1,\theta_2). 
\label{eq.symmetrydef2}
\end{equation}
Here we assume that the matrix elements ${M_{ij}}$ are 
commuting numbers. Solving the defining relation 
(\ref{eq.symmetrydef}) or (\ref{eq.symmetrydef2}), 
we have found that the symmetry matrix ${M}$ 
for the fermionic ${R}$-matrix takes the following form
\begin{equation}
M = \left(
     \begin{array}{cccc}
               M_{11} &   0    &    0    &   M_{14}   \\
                0     & M_{22} & M_{23}  &     0      \\
                0     & M_{32} & M_{33}  &     0      \\
               M_{41} &   0    &    0    &   M_{44}   
     \end{array}
    \right), \label{eq.symmetrymatrix}
\end{equation}
with the
 condition 
\begin{equation}
M_{11} M_{44} - M_{41} M_{14} = M_{22} M_{33} - M_{23} M_{32}. 
\label{eq.conditions}
\end{equation}
We denote the submatrices of ${M}$ as 
\begin{eqnarray}
& & M_{\rm charge} = \left( \begin{array}{cc}
                       M_{11} & M_{14} \\
                       M_{41} & M_{44} \\
                    \end{array}
             \right),  
M_{\rm spin} = \left( \begin{array}{cc}
                       M_{22} & M_{23} \\
                       M_{32} & M_{33} \\
                  \end{array}
           \right). \nonumber \\
\end{eqnarray}
Then the condition (\ref{eq.conditions}) can be written 
\begin{equation}
{\rm det} M_{\rm charge} = {\rm det} M_{\rm spin} \equiv \Delta(M).
\end{equation}
Since an overall constant is not relevant for the integrability, 
we put
\begin{equation}
\Delta(M) = 1. \label{eq.determinant}
\end{equation}
In other words, the submatrices ${M_{\rm charge}, M_{\rm spin}}$
belong to ${SL(2,{\bf C})}$.
The symmetry matrix ${M}$ reflects the ${SO(4)}$ symmetry of 
the Hamiltonian. It is to be remarked that the symmetry matrix 
${\bar{M}}$ of Shastry's ${R}$-matrix
${\check{R}_{12}(\theta_1,\theta_2)}$, which is defined by
\begin{equation}
\left[ \check{R}_{12}(\theta_1,\theta_2), \bar{M} \otimes \bar{M}
\right] = 0, \label{eq.Shastrysymmetrydef}
\end{equation}
 is not of the form (\ref{eq.symmetrymatrix}) 
and (\ref{eq.conditions}) \cite{Shiroishi5}. 
Shastry's ${R}$-matrix ${\check{R}_{12}(\theta_1,\theta_2)}$ is
related to the fermionic ${R}$-matrix 
${\check{\cal R}_{12}(\theta_1,\theta_2)}$ through 
the formula \cite{Olmedilla1}
\begin{equation}
\check{R}_{12}(\theta_1,\theta_2) 
= W_{12}^{-1} \check{\cal R}_{12}(\theta_1,\theta_2) W_{12}, 
\label{eq.ShastryfermiR}
 \end{equation}
where ${W_{12}}$ is a diagonal ${16 \times 16}$ matrix
\begin{equation}
W_{12} = \rm{diag}(1,1,- \rm{i},- \rm{i},- \rm{i},- \rm{i},1,1,%
-1,-1,\rm{i},\rm{i},\rm{i},\rm{i},-1,-1). \nonumber \\
\end{equation}
Since (\ref{eq.ShastryfermiR}) is not a gauge transformation, 
the symmetry of Shastry's ${R}$-matrix may be different from that 
of the fermionic ${R}$-matrix. The fermionic ${R}$-matrix is 
more suitable when we investigate the symmetry of the fermionic 
Hamiltonian in the framework of the QISM. 
 
Using the symmetry matrix, one may twist the periodic boundary
condition. The symmetry matrix ${M}$ can be inserted into the 
transfer matrix without breaking the integrability \cite{Sklyanin}
\begin{equation}
t(\theta;M) = {\rm str} \left\{ M T(\theta) \right\}. \label{eq.Mtmatrix}
\end{equation}
The generalized transfer matrix ${t(\theta;M)}$ still constitutes 
a commuting family
\begin{equation}
\left[ t(\theta_{1};M), t(\theta_{2};M) \right] = 0.
\end{equation}
The series expansion of ${t(\theta;M)}$ around ${\theta = 0}$ gives
rise to a Hamiltonian ${\hat{H}(M)}$
\begin{equation}
t(\theta;M) = {\rm str} \left\{ M T(0) \right\} 
\left( 1 + \theta {\hat{H}}(M) + \cdots \right)
\end{equation}
where
\begin{eqnarray}
\hat{H}(M) & = & - \sum_{m=1}^{N-1} \sum_{s = \uparrow \downarrow} 
(c_{m s}^{\dagger} c_{m+1 s} + c_{m+1 s}^{\dagger} c_{m s}) 
+ U \sum_{m=1}^{N} ( n_{m \uparrow} - \frac{1}{2} ) 
( n_{m \downarrow} - \frac{1}{2} ) \nonumber \\
& & - \left( M_{11} M_{22} c_{1 \uparrow}^{\dagger} + 
{\rm i} M_{11} M_{32} c_{1 \downarrow}^{\dagger} 
- M_{41} M_{32} c_{1 \uparrow} - {\rm i} M_{41} M_{22} 
c_{1 \downarrow} \right) c_{N \uparrow} \nonumber \\
& & - c_{N \uparrow}^{\dagger} \left( M_{33} M_{44} c_{1 \uparrow} + 
{\rm i} M_{44} M_{23} c_{1 \downarrow} - M_{14} M_{23} 
c_{1 \uparrow}^{\dagger} - {\rm i} M_{14} M_{33} 
c_{1 \downarrow}^{\dagger} \right) \nonumber \\
& & - \left( M_{11} M_{33} c_{1 \downarrow}^{\dagger} 
- {\rm i} M_{11} M_{23} c_{1 \uparrow}^{\dagger} - M_{41} M_{23} 
c_{1 \downarrow} + {\rm i} M_{41} M_{33} c_{1 \downarrow} \right) 
c_{N \downarrow} \nonumber \\
& & - c_{N \downarrow}^{\dagger} \left( M_{22} M_{44} c_{1 \downarrow}
- {\rm i} M_{44} M_{32} c_{1 \uparrow} - M_{14} M_{32} 
c_{1 \downarrow}^{\dagger} + {\rm i} M_{14} M_{22} c_{1 \uparrow} \right). 
\label{eq.twistedHubbard}
\end{eqnarray}
It is easy to see that the choice ${M = I}$ corresponds 
to the periodic boundary condition (\ref{eq.PBC}) \cite{Olmedilla1}.

Further, we assume the submatrices ${M_{\rm charge}}$ and 
${M_{\rm spin}}$  belong to ${SU(2)}$.  Namely,
\begin{eqnarray}
& &  M_{44} = M_{11}^{*}, \ \ M_{41} = - M_{14}^{*}, \ \ |M_{11}|^{2}
+ |M_{14}|^{2} = 1, \label{eq.MchargeSU2} \\
& &  M_{33} = M_{22}^{*}, \ \ M_{32} = - M_{23}^{*}, \ \ |M_{22}|^{2} 
+ |M_{23}|^{2} = 1. \label{eq.MspinSU2} 
\end{eqnarray}
Here the symbol * means the complex conjugation.
In this case, we find that the periodic boundary condition is twisted 
as follows
\begin{eqnarray}
c_{N+1 \uparrow}^{\dagger} & = & M_{11} M_{22} c_{1 \uparrow}^{\dagger} 
- {\rm i} M_{11} M_{23}^{*} c_{1 \downarrow}^{\dagger} 
- M_{14}^{*} M_{23}^{*} c_{1 \uparrow} + {\rm i} M_{14}^{*} M_{22} 
c_{1 \downarrow}, \nonumber \\
c_{N+1 \uparrow} & = & M_{11}^{*} M_{22}^{*} c_{1 \uparrow} + 
{\rm i} M_{11}^{*} M_{23} c_{1 \downarrow} - M_{14} M_{23} 
c_{1 \uparrow}^{\dagger} - {\rm i} M_{14} M_{22}^{*} 
c_{1 \downarrow}^{\dagger} , \nonumber \\
c_{N+1 \downarrow}^{\dagger} & = & M_{11} M_{22}^{*} 
c_{1 \downarrow}^{\dagger} 
- {\rm i} M_{11} M_{23} c_{1 \uparrow}^{\dagger} 
 + M_{14}^{*} M_{23} c_{1 \downarrow} - {\rm i} M_{14}^{*}
M_{22}^{*} c_{1 \uparrow}, \nonumber \\
c_{N+1 \downarrow} & = & M_{22} M_{11}^{*} c_{1 \downarrow} + 
{\rm i} M_{11}^{*} M_{23}^{*} c_{1 \uparrow} 
 + M_{14} M_{23}^{*} c_{1 \downarrow}^{\dagger} + 
{\rm i} M_{14} M_{22} c_{1 \uparrow}^{\dagger}. 
\label{eq.generalTPBC}
\end{eqnarray}
The relations (\ref{eq.generalTPBC}) can be expressed 
in a matrix form
\begin{eqnarray}
\left(
  \begin{array}{cc}
      c_{N+1 \downarrow}^{\dagger}  & {\rm i} c_{N+1 \uparrow} \\
      {\rm i} c_{N+1 \uparrow}^{\dagger} & c_{N+1 \downarrow}
  \end{array}
\right) & = & M_{\rm spin}^{-1} \left(
  \begin{array}{cc}
      c_{1 \downarrow}^{\dagger}  & {\rm i} c_{1 \uparrow} \\
      {\rm i} c_{1 \uparrow}^{\dagger} & c_{1 \downarrow}
  \end{array}
\right) M_{\rm charge}. \label{eq.SO(4)rotation}
\end{eqnarray}
 This shows that the periodic boundary condition can be 
rotated by the group ${SU(2) \times SU(2)}$. 
A quite similar ${SO(4)}$ rotation was observed by Affleck
\cite{Affleck}. 

Since the choices ${M_{\rm spin} =
- {\bf 1}, M_{\rm charge} = {\bf 1}}$ and ${M_{\rm spin} = {\bf 1},
M_{\rm charge} = - {\bf 1}}$ induce the same transformation
(\ref{eq.SO(4)rotation}), the exact group symmetry is 
${SO(4) = [SU(2) \times SU(2)] / {\bf Z}_{2}}$. 

Now let us consider the diagonal ${M}$ parametrized as follows, 
\begin{eqnarray}
M&=&{\rm e}^{{\rm i} \frac{\phi}{2} \sigma^{z}} \otimes 
{\rm e}^{{\rm i} \frac{\psi}{2} \sigma^{z}} \nonumber \\
&=&\left(
     \begin{array}{cccc}
                {\rm e}^{{\rm i}\frac{\phi+\psi}{2}}&0&0&0 \\
                0&{\rm e}^{{\rm i}\frac{\phi-\psi}{2}}&0&0   \\
                0&0&{\rm e}^{-{\rm i}\frac{\phi-\psi}{2}}&0   \\
                0&0&0&{\rm e}^{-{\rm i}\frac{\phi+\psi}{2}}      
     \end{array}
    \right). \label{eq.diagonalTPBC}
\end{eqnarray}
From (\ref{eq.generalTPBC}), the corresponding twisted periodic
boundary condition is 
\begin{eqnarray}
& & c_{N+1 \uparrow}^{\dagger} = {\rm e}^{{\rm i} \phi} 
c_{1 \uparrow}^{\dagger}, \ \ c_{N+1 \uparrow} 
= {\rm e}^{- {\rm i} \phi} c_{1 \uparrow}, \nonumber \\
& & c_{N+1 \downarrow}^{\dagger} = {\rm e}^{{\rm i} \psi} 
c_{1 \downarrow}^{\dagger}, \ \  c_{N+1 \downarrow} 
= {\rm e}^{ - {\rm i} \psi} c_{1 \downarrow}, \label{eq.TwistedPBC}
\end{eqnarray}
which is identical to (\ref{eq.TPBC}).  
Especially, if we take ${\phi = \pi}$ and ${\psi = -\pi}$, 
(\ref{eq.diagonalTPBC}) becomes
\begin{equation}
M = \left( \begin{array}{cccc}
                   1 & 0  & 0 & 0 \\
                   0 & -1 & 0 & 0 \\
                   0 & 0 & -1 & 0 \\
                   0 & 0 & 0  & 1 
           \end{array}
    \right) =  \sigma^{z} \otimes \sigma^{z}.
\end{equation}
In this case, the supertrace in the transfer matrix
(\ref{eq.Mtmatrix}) becomes the usual trace
\begin{equation}
{\rm str} \left\{ \left( \sigma^{z} \otimes \sigma^{z} \right) 
T(\theta) \right\} = {\rm tr} T(\theta).
\end{equation}
Thus, as remarked in ref. 23, we can take the trace of the monodromy 
matrix instead of the supertrace. 
However, the boundary condition in this case should be anti-periodic 
\begin{eqnarray}
& & c_{N+1 \uparrow}^{\dagger} = - c_{1 \uparrow}^{\dagger}, \ \ 
c_{N+1 \uparrow} = - c_{1 \uparrow}, \nonumber \\
& & c_{N+1 \downarrow}^{\dagger} = - c_{1 \downarrow}^{\dagger}, \ \  
c_{N+1 \downarrow} = - c_{1 \downarrow}. \label{eq.AntitwistedPBC}
\end{eqnarray}

Next we consider a discrete symmetry related to the partial
particle-hole transformation (\ref{eq.particlehole}).  
We solve the following equation 
\begin{equation}
{\cal R}_{12}(\theta_1,\theta_2;U) ( N \sotimes I )( I \sotimes N ) 
\nonumber \\
= ( I \sotimes N )( N \sotimes I ) {\cal R}_{12}(\theta_1,\theta_2;-U). 
\label{eq.discrete}
\end{equation}
with a constant matrix ${N=(N_{ij})}$. 
Here we explicitly write the ${U}$-dependmence of 
the fermionic ${R}$-matrix. Note that the coupling constant of 
the fermionic ${R}$-matrix in the RHS is ${- U}$, or equivalently 
${h_{1} \rightarrow - h_1, h_{2} \rightarrow - h_{2}}$. 
By solving the defining realtion (\ref{eq.discrete}), 
we find that the constant matrix ${N}$ has a form, 
\begin{equation}
N = \left(
     \begin{array}{cccc}
                0      & N_{12}  & N_{13}    &     0      \\
                N_{21} &   0     &   0       &   N_{24}   \\
                N_{31} &   0     &   0       &   N_{34}   \\
                0      & N_{42}  &  N_{43}   &     0      
     \end{array}
    \right), \label{eq.discretematrix}
\end{equation}
with a condition
\begin{equation}
\Delta(N) = N_{12} N_{43} - N_{13} N_{42} 
= N_{21} N_{34} - N_{31} N_{24}.
\end{equation}
We set 
\begin{equation}
\Delta(N) = 1,
\end{equation}
as before.
The matrix ${N}$ can be written as
\begin{equation}
N =  \left( {\bf 1} \otimes \sigma^{x} \right)
    \left(
     \begin{array}{cccc}
                N_{21} &   0     &   0       &   N_{24}   \\
                0      & N_{12}  & N_{13}    &     0      \\
                0      & N_{42}  &  N_{43}   &     0      \\
                N_{31} &   0     &   0       &   N_{34}   
    \end{array}
    \right). 
\end{equation}
Here and hereafter, ${\bf 1}$ means the ${2 \times 2}$ identity matrix.
Thus  ${N}$ is a composition of the symmetry matrix 
(\ref{eq.symmetrymatrix}) and 
\begin{equation}
Q = Q^{-1} = {\bf 1} \otimes \sigma^{x} = \left(
                         \begin{array}{cccc}
                           0 & 1 & 0 & 0 \\
                           1 & 0 & 0 & 0 \\
                           0 & 0 & 0 & 1 \\
                           0 & 0 & 1 & 0 
                          \end{array}
                     \right). \label{eq.Q}
\end{equation}  
In terms of this matrix, the fermionic ${R}$-matrix with the coupling 
constant ${U}$ is transformed into the fermionic ${R}$-matrix 
with the coupling constant ${-U}$ as follows
\begin{equation}
{\stackrel{1}{Q^{-1}}} {\stackrel{2}{Q^{-1}}} 
{\cal R}_{12}(\theta_1,\theta_2;U) \stackrel{1}{Q} \stackrel{2}{Q} 
= {\cal R}_{12}(\theta_1,\theta_2;-U). \label{eq.Rdiscrete}
\end{equation}
The matrix ${Q}$ is related to the partial particle-hole
transformation (\ref{eq.particlehole}) of the Hamiltonian. 
There are other choices for the matrix ${Q}$. For example,
\begin{equation}
Q =  \sigma^{z} \otimes \sigma^{y}  = \left(
                         \begin{array}{cccc}
                           0 & -{\rm i} & 0 & 0 \\
                          {\rm i} & 0 & 0 & 0 \\
                           0 & 0 & 0 & {\rm i} \\
                           0 & 0 & -{\rm i} & 0 
                          \end{array}
                     \right)
\end{equation}  
also satisfies (\ref{eq.Rdiscrete}), which may be more appropriate 
as the partial particle-hole transformation 
(\ref{eq.particlehole}) \cite{Gohmann}. 
However, for simplicity, we choose the matrix (\ref{eq.Q}) as the
partial particle-hole transformation of the fermionic ${R}$-matrix. 

\vspace{20pt}
\newpage
\begin{flushleft}
{\large \bf \S 4. Graded Reflection Equations for the Fermionic 
${\mbox{\boldmath $R$}}$-Matrix}
\end{flushleft}
\setcounter{equation}{0}
\renewcommand{\theequation}{4.\arabic{equation}}
In this section  we investigate the integrability of the 1D Hubbard
model with open boundary condition in terms of the fermionic
${R}$-matrix ${{\cal R}_{12}(\theta_1,\theta_2)}$.
We introduce an associative algebra ${{\cal{T}}_{-}}$ defined by the 
fermionic ${R}$-matrix \cite{Sklyanin,Mezincescu,Kulish2}, 
\begin{eqnarray}
& & {\cal R}_{12}(\theta_1,\theta_2) 
{\stackrel{1}{\cal{T}}}_{-}(\theta_1)
{\cal R}_{21}(\theta_2,-\theta_1)
{\stackrel{2}{\cal{T}}}_{-}(\theta_2) \nonumber \\ 
& & = {\stackrel{2}{\cal{T}}}_{-}(\theta_2) 
{\cal R}_{12}(\theta_1,-\theta_2)
{\stackrel{1}{\cal{T}}}_{-}(\theta_1) 
{\cal R}_{21}(-\theta_2,-\theta_1), 
\label{eq.RET1}
\end{eqnarray}
where
\begin{equation}
{\stackrel{1}{\cal{T}}}_{-}(\theta_1) \equiv {\cal{T}}_{-}(\theta_1) 
\sotimes I, \ \ \ \ {\stackrel{2}{\cal{T}}}_{-}(\theta_2) \equiv I 
\sotimes {\cal{T}}_{-}(\theta_2).
\end{equation}  
The relation (\ref{eq.RET1}) is called the graded reflection 
equation (graded RE). 

The following theorem is fundamental for the application of the
associative algebra ${{\cal T}_{-}}$. \\ 
{{\it Theorem} \cite{Sklyanin,Kulish2}} \\
Let ${{\tilde{\cal T}}_{-}(\theta)}$ be some representation of 
the associative algebra ${{\cal T}_{-}}$ (\ref{eq.RET1}) and
${T(\theta)}$ of the associative algebra ${T}$ (\ref{eq.globalGYBR2}).
Then ${{\cal T}_{-}(\theta)}$ defined by
\begin{equation}
{\cal T}_{-}(\theta) 
= T(\theta) \tilde{\cal T}_{-}(\theta) T^{-1}(-\theta), 
\label{eq.covariance}
\end{equation}
is also a representation of ${{\cal T}_{-}}$ provided that the matrix 
elements of ${\tilde{\cal T}_{-}(\theta)}$ and ${T(\theta)}$ commute. 
This property is sometimes called the covariance property of the
(graded) reflection equation \cite{Kulish2}.
For the application to the 1D Hubbard model, we choose ${T(\theta)}$
and ${\tilde{\cal T}_{-}(\theta)}$ as
\begin{eqnarray}
T(\theta) &=&{\cal L}_{N}(\theta) {\cal L}_{N-1}(\theta) \cdots 
{\cal L}_{1}(\theta), \nonumber \\
 \tilde{\cal T}_{-}(\theta) & = & K_{-}(\theta).
\end{eqnarray}
Here ${K_{-}(\theta)}$ is a constant supermatrix satisfying 
the graded RE (\ref{eq.RET1}), 
\begin{eqnarray}
& & {\cal R}_{12}(\theta_1,\theta_2) 
{\stackrel{1}{K}}_{-}(\theta_1) 
{\cal R}_{21}(\theta_2,-\theta_1) 
{\stackrel{2}{K}}_{-}(\theta_2) \nonumber \\ 
& & = {\stackrel{2}{K}}_{-}(\theta_2) 
{\cal R}_{12}(\theta_1,-\theta_2) 
{\stackrel{1}{K}}_{-}(\theta_1) 
{\cal R}_{21}(-\theta_2, -\theta_1).  
\label{eq.REK1}
\end{eqnarray}
The ${K}$-matrix ${K_{-}(\theta)}$ specifies the integrable boundary 
condition. In this paper we assume that the matrix elements of 
${K_{-}(\theta)}$ are commuting numbers and only the ${\it even}$
elements with respect to the parity (\ref{eq.parity}) 
are non zero \cite{Mezincescu},
\begin{equation}
K_{-}(\theta) 
= \left( \begin{array}{cccc}
      K_{11}(\theta) & 0 & 0 & K_{14}(\theta) \nonumber \\
       0 & K_{22}(\theta) & K_{23}(\theta) & 0   \nonumber \\
       0 & K_{32}(\theta) & K_{33}(\theta) & 0   \nonumber \\
        K_{41}(\theta) & 0 & 0 &  K_{44}(\theta) 
           \end{array}
     \right). \label{eq.condK}
\end{equation}
To construct a transfer matrix, we introduce another ${K}$-matrix 
${K_{+}(\theta)}$ satisfying the conjugated graded RE
\begin{eqnarray}
& &  {\cal R}_{21}^{\rm st_1,\overline{\rm st}_2}(\theta_2,\theta_1) 
\stackrel{1\;\;\;\;}{K_{+}^{\rm st_1}}(\theta_1) 
{\cal R}_{12}^{\rm st_1,\overline{\rm st}_2}(- \theta_1,\theta_2) 
\stackrel{2\;\;\;\;}{K_{+}^{\overline{\rm st}_2}}(\theta_2) \nonumber \\
& & = \stackrel{2\;\;\;\;}{K_{+}^{\overline{\rm st}_2}}(\theta_2) 
{\cal R}_{21}^{\rm st_1,\overline{\rm st}_2}(-\theta_2, \theta_1) 
\stackrel{1\;\;\;\;}{K_{+}^{\rm st_1}}(\theta_1) 
{\cal R}_{12}^{\rm st_1,\overline{\rm st}_2}(-\theta_1,-\theta_2). 
\label{eq.REK2}
\end{eqnarray}
We assume that the ${K}$-matrix ${K_{+}(\theta)}$ is also of the form 
(\ref{eq.condK}).  Then the supertransposition to the ${K}$-matrix 
${K_{+}(\theta)}$ reduce to the usual transposition. 

Now we define the transfer matrix by
\begin{eqnarray}
\tau(\theta) & = & {\rm str} \left\{ K_{+}(\theta) {\cal T}_{-}(\theta) 
\right\} \nonumber \\
& = & {\rm str} \left\{ K_{+}(\theta) T(\theta) K_{-}(\theta) 
T^{-1}(-\theta) \right\}. \label{eq.transfermatrix}
\end{eqnarray} 
Using the unitarity (\ref{eq.unitarity}), the crossing unitarity 
(\ref{eq.crossingunitarity}) and the graded REs (\ref{eq.RET1}), 
(\ref{eq.REK2}), we can show that the transfer matrix ${\tau(\theta)}$ 
constitutes a commutative family \cite{Sklyanin}, i.e.,
\begin{equation}
\left[ \tau(\theta_1), \tau(\theta_2) \right] = 0,
\end{equation}
which shows the existence of the mutually commuting conserved currents 
including the Hamiltonian.

We note
\begin{equation}
T^{-1}(-\theta) = {\cal{L}}_{1}^{-1} (- \theta) \cdots 
{\cal{L}}_{N}^{-1} (- \theta),
\end{equation}
where
\begin{eqnarray}
& & {\cal{L}}_{m}^{-1}(- \theta) \nonumber \\ 
& & =  \frac{1}{\cos^4 \theta}
\left(
 \begin{array}{cccc}
 - {\rm e}^{h} f_{m \uparrow}^{*}(\theta) 
   f_{m \downarrow}^{*}(\theta) &  
   {\rm i} f_{m \uparrow}^{*}(\theta) c_{m \downarrow} &
   c_{m \uparrow} f_{m \downarrow}^{*}(\theta) &
 - {\rm i} {\rm e}^{h} c_{m \uparrow} c_{m \downarrow} \\
 - f_{m \uparrow}^{*}(\theta) c_{m \downarrow}^{\dagger} &
   {\rm e}^{-h} f_{m \uparrow}^{*}(\theta) 
    g_{m \downarrow}^{*}(\theta) &
  - {\rm e}^{-h} c_{m \uparrow} c_{m \downarrow}^{\dagger} &
  c_{m \uparrow} g_{m \downarrow}^{*}(\theta) \\
  - {\rm i} c_{m \uparrow}^{\dagger} f_{m \downarrow}^{*}(\theta) &
  - {\rm e}^{-h} c_{m \uparrow}^{\dagger} c_{m \downarrow} &
  {\rm e}^{-h} g_{m \uparrow}^{*}(\theta) 
   f_{m \downarrow}^{*}(\theta) &
  - {\rm i} g_{m \uparrow}^{*}(\theta) c_{m \downarrow} \\
  {\rm i} {\rm e}^{h} c_{m \uparrow}^{\dagger} 
   c_{m \downarrow}^{\dagger} &
  - {\rm i} c_{m \uparrow}^{\dagger} g_{m \downarrow}^{*}(\theta) &
  g_{m \uparrow}^{*}(\theta) c_{m \downarrow}^{\dagger} &
  - {\rm e}^{h} g_{m \uparrow}^{*}(\theta) 
   g_{m \downarrow}^{*}(\theta)
 \end{array}
\right) \nonumber \\
& & = \frac{1}{\cos^4 \theta} {\cal L}(\theta) 
\left( {\bar{\cal L}}_{m \uparrow}(\theta) 
\sotimes {\bar{\cal L}}_{m \downarrow}(\theta) \right) {\cal L}(\theta). 
\end{eqnarray}
Here we define
\begin{equation}
{\bar{\cal{L}}}_{m \uparrow}(\theta) 
= \left(
     \begin{array}{cc}
    - f_{m \uparrow}^{*}(\theta) &  c_{m \uparrow} \\
    - {\rm i} c_{m \uparrow}^{\dagger} & g_{m \uparrow}^{*}(\theta) \\
                    \end{array}
                    \right), \ \ \ \ 
{\bar{\cal{L}}}_{m \downarrow}(\theta) 
= \left(
      \begin{array}{cc}
       f_{m \downarrow}^{*}(\theta) & - {\rm i} c_{m \downarrow} \\
       c_{m \downarrow}^{\dagger} & - g_{m \downarrow}^{*}(\theta)  
       \end{array}
    \right).
\end{equation}

\vspace{20pt}
\begin{flushleft}
{\large \bf \S 5. Integrability of the 1D Hubbard Model %
with Boundary Fields}
\end{flushleft}
\setcounter{equation}{0}
\renewcommand{\theequation}{5.\arabic{equation}}
In this section we solve the graded REs (\ref{eq.REK1}) 
and (\ref{eq.REK2}) for ${K_{-}(\theta)}$ and ${K_{+}(\theta)}$. 
First, we note a relation between the ${R}$-matrix
\begin{equation}
R_{12}(\theta_1,\theta_2) = R_{21}^{*}(-\theta_2,-\theta_1).
\end{equation}
Then the graded RE (\ref{eq.REK1}) for  ${K_{-}(\theta)}$ is cast into
a form 
\begin{eqnarray}
& &  {\cal R}_{12}(\theta_1,\theta_2) 
{\stackrel{1}{K}}_{-}(\theta_1) 
{\cal R}_{12}^{*}(\theta_1,-\theta_2) 
{\stackrel{2}{K}}_{-}(\theta_2) \nonumber \\ 
& &  = {\stackrel{2}{K}}_{-}(\theta_2) 
{\cal R}_{12}(\theta_1,-\theta_2) 
{\stackrel{1}{K}}_{-}(\theta_1) 
{\cal R}_{12}^{*}(\theta_1,\theta_2), \label{eq.REK11}
\end{eqnarray}
with the graded tensor products,
\begin{equation}
{\stackrel{1}{K}}_{-}(\theta_1) \equiv K_{-}(\theta_1) \sotimes I, \ \
\ \ 
{\stackrel{2}{K}}_{-}(\theta_2) \equiv I \sotimes K_{-}(\theta_2).
\end{equation}  

Assuming ${K_{-}(\theta)}$ in a diagonal form
\begin{equation}
K_{-}(\theta) 
= \left(
      \begin{array}{cccc}
        x_{1}(\theta) & 0        & 0        & 0        \\
          0      & x_{2}(\theta) & 0        & 0        \\
          0      & 0        & x_{3}(\theta) & 0        \\
          0      & 0        & 0        & x_{4}(\theta) 
                   \end{array}
           \right),
\end{equation}
and substituting it into the graded RE (\ref{eq.REK11}), 
we have 10 non-trivial functional equations for 
${x_{i}(\theta), (i=1,\cdots 4)}$. 
By solving these functional equations (see Appendix), 
we obtain the following two sets of solutions 
for the diagonal ${K_{-}(\theta)}$. \\
a) ${K_{-}(\theta) = K_{-}^{(a)}(\theta;p_{-})}$:
\begin{eqnarray}
x_{1}(\theta) & = & \left(1 - p_{-} {\rm e}^{-2h} \tan \theta \right) 
\left( 1  - p_{-} {\rm e}^{2h} \tan \theta \right), \nonumber \\
x_{2}(\theta) & = & x_{3}(\theta) 
                =  {\rm e}^{-2h} \left(1 + p_{-} {\rm e}^{2h} 
\tan \theta \right) \left( 1 - p_{-} {\rm e}^{2h} \tan \theta \right), 
\nonumber \\
x_{4}(\theta) & = & \left(1 + p_{-} {\rm e}^{2h} \tan \theta \right) 
\left( 1 + p_{-} {\rm e}^{-2h} \tan \theta \right). 
\label{eq.Kminusa}
\end{eqnarray}
b) ${K_{-}(\theta) = K_{-}^{(b)}(\theta;p_{-})}$:
\begin{eqnarray}
x_{1}(\theta) & = & x_{4}(\theta)
                =  {\rm e}^{2h} \left(1 + p_{-} {\rm e}^{-2h} 
\tan \theta \right) \left( 1 - p_{-} {\rm e}^{-2h} \tan \theta
\right), \nonumber \\
x_{2}(\theta) & = & \left( 1 - p_{-} {\rm e}^{-2h} \tan \theta \right) 
\left( 1  - p_{-} {\rm e}^{2h} \tan \theta \right), \nonumber \\
x_{3}(\theta) & = & \left(1 + p_{-} {\rm e}^{-2h} \tan \theta \right) 
\left( 1 + p_{-} {\rm e}^{2h} \tan \theta \right). 
\label{eq.Kminusb}
\end{eqnarray}
Here ${p_{-}}$ is a constant parameter corresponding 
to the boundary field. Recall that the parameter ${h}$ is 
regarded as a function of the spectral parameter ${\theta}$ 
through the constraint
\begin{equation}
\frac{\sinh 2h}{\sin 2 \theta} = \frac{U}{4}.
\end{equation}
When we put ${p_{-} = 0}$, the solutions ${K_{-}^{(a)}(\theta;p_{-})}$ 
and ${K_{-}^{(b)}(\theta;p_{-})}$ are proportional to 
${{\cal L}^{2}(\theta)}$,
\begin{eqnarray}
K_{-}^{(a)}(\theta;p_{-} = 0) & = & {\rm e}^{-h} {\cal L}^{2}(\theta), \\ 
K_{-}^{(b)}(\theta;p_{-} = 0) & = & {\rm e}^{h} {\cal L}^{2}(\theta),
\end{eqnarray}
which supply the missing interaction term at the boundary site
${m=1}$. 
To see the contributions of the boundary fields clearly, we redefine
\begin{eqnarray}
\tilde{K}_{-}^{(a)}(\theta;p_{-}) 
& = & {\rm e}^{h} {\cal L}^{-1}(\theta) K_{-}^{(a)}(\theta;p_{-}) 
{\cal L}^{-1}(\theta), \label{eq.K_a} \\
\tilde{K}_{-}^{(b)}(\theta;p_{-}) 
& = & {\rm e}^{-h} {\cal L}^{-1}(\theta) K_{-}^{(b)}(\theta;p_{-}) 
{\cal L}^{-1}(\theta). \label{eq.K_b}
\end{eqnarray}
Expanding ${\tilde{K}_{-}^{(a)}(\theta;p_{-})}$ around ${\theta = 0}$, 
we have
\begin{eqnarray}
& & \tilde{K}_{-}^{(a)}(\theta;p_{-}) 
= {\bf 1} \otimes {\bf 1} - p_{-} \left( \sigma^{z} \otimes {\bf 1}+ 
{\bf 1} \otimes \sigma^{z} \right) \theta + \cdots. 
\label{eq.Kaexpansion} 
\end{eqnarray}
Now the linear term with respect to the spectral parameter produces 
the boundary chemical potential at the site ${m=1}$
\begin{equation}
-p_{-}(n_{1 \uparrow} + n_{1 \downarrow} - 1). 
\label{eq.bdrychemicalterm}
\end{equation}
Similarly, we have
\begin{eqnarray}
& & \tilde{K}_{-}^{(b)}(\theta;p_{-}) = {\bf 1} \otimes {\bf 1} - 
p_{-} \left( \sigma^{z} \otimes {\bf 1} - {\bf 1} \otimes \sigma^{z}
\right) \theta + \cdots. \label{eq.Kbexpansion}
\end{eqnarray}
The linear term in (\ref{eq.Kbexpansion}) gives the boundary 
magnetic field at the site ${m=1}$
\begin{equation}
-p_{-}(n_{1 \uparrow} - n_{1 \downarrow}). 
\label{eq.bdrymagneticterm}
\end{equation}
Thus we have shown that ${K_{-}^{(a)}(\theta;p_{-})}$ corresponds to 
the boundary chemical potential, while ${K_{-}^{(b)}(\theta;p_{-})}$ 
corresponds to the boundary magnetic field.

A comment is in order on the derivations of
(\ref{eq.bdrychemicalterm}) and (\ref{eq.bdrymagneticterm}). 
Rigorously speaking, (\ref{eq.Kaexpansion}) and 
(\ref{eq.Kbexpansion}) respectively give twice of 
(\ref{eq.bdrychemicalterm}) and (\ref{eq.bdrymagneticterm}). 
However, the bulk Hamiltonian appears twice in the expansion of 
the transfer matrix. Therefore the Hamiltonian contains 
the boundary fields as given in  (\ref{eq.bdrychemicalterm}) and 
(\ref{eq.bdrymagneticterm}).

We also remark that the diagonal solutions (\ref{eq.Kminusa}) and 
(\ref{eq.Kminusb}) coincide with those found for the coupled spin
model (\ref{eq.BoundaryCoupledSpin}) \cite{Zhou,Guan}. 
Actually, if we formulate the reflection equation in terms of
Shastry's ${R}$-matrix \cite{Zhou}, we obtain the same functional 
equations as (\ref{eq.b1})--(\ref{eq.b10}) (see Appendix) for the 
diagonal ${K_{-}(\theta)}$.

Solving the graded RE (\ref{eq.REK2}) for the diagonal
${K_{+}(\theta)}$ in a similar way, we obtain two solutions for 
the ${K}$-matrix ${K_{+}(\theta)}$ as follow.\\
a) ${K_{+}(\theta) = K_{+}^{(a)}(\theta;p_{+})}$:
\begin{eqnarray}
x_{1}(\theta) & = & \left( p_{+} + {\rm e}^{2h} 
\tan \theta \right) \left( p_{+} + {\rm e}^{-2h} 
\tan \theta \right), \nonumber \\
x_{2}(\theta) & = & x_{3}(\theta) 
=  {\rm e}^{2h} \left( p_{+} + {\rm e}^{-2h} 
\tan \theta \right) \left( p_{+} - {\rm e}^{-2h} \tan \theta \right) 
\nonumber \\
x_{4}(\theta) & = & \left(p_{+} - {\rm e}^{2h} \tan \theta \right) 
\left( p_{+} - {\rm e}^{-2h} \tan \theta \right). \label{eq.Kplusa}
\end{eqnarray}
b) ${ K_{+}(\theta) = K_{+}^{(b)}(\theta;p_{+})}$:
\begin{eqnarray}
x_{1}(\theta) & = & x_{4}(\theta) 
                =  {\rm e}^{-2h} \left(p_{+} + {\rm e}^{2h} 
\tan \theta \right) \left( p_{+} - {\rm e}^{2h} \tan \theta \right), 
\nonumber \\
x_{2}(\theta) & = & \left( p_{+} + {\rm e}^{2h} \tan \theta \right) 
\left( p_{+} + {\rm e}^{-2h} \tan \theta \right), \nonumber \\
x_{3}(\theta) & = & \left( p_{+} - {\rm e}^{2h} \tan \theta \right) 
\left(p_{+} - {\rm e}^{-2h} \tan \theta \right). \label{eq.Kplusb}
\end{eqnarray}
In (\ref{eq.Kplusa}) and (\ref{eq.Kplusb}), the parameter ${p_{+}}$ 
represents the strength of the boundary field at the site ${m=N}$.
The solution ${K_{+}^{(a)}(\theta;p_{+})}$ corresponds to the boundary 
chemical potential and  ${K_{+}^{(b)}(\theta;p_{+})}$ corresponds 
to the boundary magnetic field. 

As in the case of (\ref{eq.K_a}) and (\ref{eq.K_b}), we may redefine
\begin{eqnarray}
\tilde{K}_{+}^{(a)}(\theta;p_{+}) 
& = & {\rm e}^{-h} {\cal L}(\theta) K_{+}^{(a)}(\theta;p_{+}) 
{\cal L}(\theta), \\
\tilde{K}_{+}^{(b)}(\theta;p_{+}) 
& = & {\rm e}^{h} {\cal L}(\theta) K_{+}^{(b)}(\theta;p_{+}) 
{\cal L}(\theta).
\end{eqnarray}
Then around ${\theta = 0}$, we have
\begin{eqnarray}
\tilde{K}_{+}^{(a)}(\theta;p_{+}) 
& = & p_{+}^{2} \left( {\bf 1} \otimes {\bf 1} \right) + p_{+} \left
( \sigma^{z} \otimes {\bf 1} + {\bf 1} \otimes \sigma^{z} \right)
\theta + \left( \sigma^{z} \otimes \sigma^{z} \right) \theta^2 
+ \cdots, \label{eq.K+aexpand} \\
\tilde{K}_{+}^{(b)}(\theta;p_{+}) & = & p_{+}^{2} \left( {\bf 1} 
\otimes {\bf 1} \right) + p_{+} \left( \sigma^{z} \otimes {\bf 1} 
- {\bf 1} \otimes \sigma^{z} \right) \theta 
- \left( \sigma^{z} \otimes \sigma^{z} \right) \theta^2 + \cdots. 
\label{eq.K+bexpand}
\end{eqnarray}
From (\ref{eq.K+aexpand}) and (\ref{eq.K+bexpand}), we find 
the following properties
\begin{equation}
{\rm str} \tilde{K}_{+}^{(a)}(\theta=0;p_{+}) 
={\rm str} \tilde{K}_{+}^{(b)}(\theta=0;p_{+}) =0,\label{eq.prop1} 
\end{equation}
\begin{equation}
{\rm str} \left( \frac{\rm d}{\rm d \theta} \tilde{K}_{+}^{(a)}
(\theta;p_{+}) \Big|_{\theta=0} \right)  
= {\rm str} \left( \frac{\rm d}{\rm d \theta} \tilde{K}_{+}^{(b)}
(\theta;p_{+}) \Big|_{\theta=0} \right) = 0. \label{eq.prop2}
\end{equation}

Using the solutions (\ref{eq.Kminusa}), (\ref{eq.Kminusb}), 
(\ref{eq.Kplusa}) and (\ref{eq.Kplusb}), we can construct the four 
integrable transfer matrices ${\tau(\theta) =}$ 
${{\rm str} \left\{ K_{+}(\theta) T(\theta) %
K_{-}(\theta) T^{-1}(-\theta) \right\} }$. 
\begin{eqnarray} 
& & {\rm Case \ A}. \ \ K_{-}(\theta) 
= K_{-}^{(a)}(\theta;p_{1}), 
\ \ K_{+}(\theta) = K_{+}^{(a)}(\theta;p_{N}). \nonumber \\
& & {\rm Case \ B}. \ \ K_{-}(\theta) 
= K_{-}^{(b)}(\theta;p_{1}), 
\ \ K_{+}(\theta) = K_{+}^{(b)}(\theta;p_{N}). \nonumber \\
& & {\rm Case \ C}. \ \ K_{-}(\theta) 
= K_{-}^{(b)}(\theta;p_{1}), 
\ \  K_{+}(\theta) = K_{+}^{(a)}(\theta;p_{N}). \nonumber \\
& & {\rm Case \ D}. \ \ K_{-}(\theta) 
= K_{-}^{(a)}(\theta;p_{1}), 
\ \ K_{+}(\theta) = K_{+}^{(b)}(\theta;p_{N}). 
\end{eqnarray}
These results are consistent with the coordinate Bethe ansatz 
for the cases (\ref{eq.4cases}). 
The Hamiltonian (\ref{eq.BoundaryHubbard}) is obtained 
by expanding the transfer matrix with respect to the spectral 
parameter ${\theta}$
\begin{equation}
\tau(\theta) = C_{1} \theta + C_{2} \theta^2 + 
C_{3} \left( {\hat{H}}_{\rm open} + {\rm const.} \right) \theta^3
 + \cdots. \label{eq.transferhamiltonian}
\end{equation}
Here ${C_{i} \ \ (i=1,2,3,\cdots)}$ are some scalar functions. 
We omit the detailed derivation of (\ref{eq.transferhamiltonian}) 
in this paper, which can be carried out using the method in ref. 22. 
Due to the properties (\ref{eq.prop1}) and (\ref{eq.prop2}), 
the Hamiltonian ${{\hat{H}}_{\rm open}}$ appears at the third order 
of the expansion. 

The commutativity of the transfer matrix  
\begin{equation}
\left[ \tau(\theta_{1}), \tau(\theta_{2}) \right] = 0
\end{equation}
ensures the existence of an infinite number of the conserved currents 
in involution. The exact integrability of the 1D Hubbard model 
with the boundary fields (\ref{eq.BoundaryHubbard}) is established 
in this way.

The ${K}$-matrices (\ref{eq.Kplusa}) and (\ref{eq.Kplusb}) are
slightly different from those for Shastry's
${R}$-matrix \cite{Zhou,Guan}, which we denote
${\bar{K}_{+}(\theta)}$. The former can be obtained from the latter 
by multiplying the diagonal matrix ${\sigma^{z} \otimes \sigma^{z}}$, 
\begin{equation}
K_{+}(\theta) = \left( \sigma^{z} \otimes \sigma^{z} \right) 
\bar{K}_{+}(\theta).
\end{equation}
Due to the definition of the supertrace (\ref{eq.supertrace}), 
both formulations provide the same transfer matrix
\begin{equation}
{\rm str} \left\{ K_{+}(\theta) T(\theta) 
K_{-}(\theta) T^{-1}(-\theta) \right\} 
= {\rm tr} \left\{ \bar{K}_{+}(\theta) T(\theta) 
K_{-}(\theta) T^{-1}(-\theta) \right\}.
\end{equation}

\vspace{20pt}
\newpage
\begin{flushleft}
{\large \bf \S 6. Non-Diagonal Solutions of %
the Graded Reflection Equation}
\end{flushleft}
\setcounter{equation}{0}
\renewcommand{\theequation}{6.\arabic{equation}}
 We can construct non-diagonal solutions of the graded RE 
(\ref{eq.REK1}) by use of the covariance property 
(\ref{eq.covariance}) and the symmetry matrix ${M}$. 
For example, we have a solution
\begin{eqnarray}
 K_{-}^{(a)}(\theta;p_{-},M_{\rm charge}) 
 & = & M K^{(a)}_{-}(\theta;p_{-}) M^{-1} \nonumber \\
 & = & \left(
      \begin{array}{cccc}
      K_{11}(\theta) &  0      &  0     & K_{14}(\theta) \\
      0      &  K_{22}(\theta) &  0     & 0 \\
      0      &  0      & K_{33}(\theta) & 0 \\
      K_{41}(\theta) &  0      &   0    & K_{44}(\theta) 
      \end{array}
      \right),  \label{eq.generalKminusa}
\end{eqnarray}  
where
\begin{eqnarray}
K_{11}(\theta) & = & 1 - 2 p_{-} (M_{11} M_{44} + M_{14} M_{41}) 
\cosh2h \tan \theta + p_{-}^2 \tan^2 \theta , \nonumber \\
K_{22}(\theta) & = & K_{33}(\theta) = e^{-2h} \left( 1 - p_{-}^{2} 
e^{4h} \tan^2 \theta \right), \nonumber \\
K_{44}(\theta) & = & 1 + 2 p_{-} (M_{11} M_{44} + M_{14} M_{41}) 
\cosh 2h \tan \theta + p_{-}^2 \tan^2 \theta , \nonumber \\
K_{14}(\theta) & = & 4 p_{-} M_{11} M_{14} \cosh 2h \tan \theta, 
\nonumber \\
K_{41}(\theta) & = & - 4 p_{-} M_{41} M_{44} \cosh 2h \tan \theta. 
\label{eq.generalsola}
\end{eqnarray} 
Note that ${K_{-}^{(a)}(\theta;p_{-},M_{\rm charge})}$ does not 
depend on the submatrix ${M_{\rm spin}}$.

Since the relation
\begin{eqnarray}
M_{11} M_{44} + M_{14} M_{41} 
& = & \left\{ \left( M_{11} M_{44} - M_{14} M_{41} \right)^2 
+ 4 M_{11} M_{14} M_{41} M_{44} \right\}^{\frac{1}{2}} \nonumber \\
& = & \left( 1 + 4 M_{11} M_{14} M_{41} M_{44} \right)^{\frac{1}{2}}
\end{eqnarray}
holds, the solution (\ref{eq.generalsola}) depends on three arbitrary 
parameters, 
\begin{equation}
 p_{-}, \ \ \ \  \alpha = M_{11} M_{14}, \ \ \ \ 
\beta = - M_{41} M_{44}.
\end{equation}
The corresponding boundary term is
\begin{eqnarray}
& & - p_{-} \big\{ (M_{11} M_{44} + M_{14} M_{41}) 
\left( n_{1 \uparrow} + n_{1 \downarrow} - 1 \right) 
 - 2 {\rm i} M_{11} M_{14} c_{1 \uparrow}^{\dagger} 
c_{1 \downarrow}^{\dagger} - 2 {\rm i} M_{41} M_{44} c_{1 \downarrow} 
c_{1 \uparrow} \big\} \nonumber \\
& & = - p_{-} \big\{ (1 - 4 \alpha \beta)^{\frac{1}{2}} 
\left( n_{1 \uparrow} + n_{1 \downarrow} - 1 \right) 
 - 2 {\rm i} \alpha c_{1 \uparrow}^{\dagger} c_{1 \downarrow}^{\dagger}  
+  2 {\rm i} \beta c_{1 \downarrow} c_{1 \uparrow} \big\}. 
\label{eq.boundaryterma}
\end{eqnarray}
Note the existence of the term ${c_{1 \uparrow}^{\dagger} 
c_{1 \downarrow}^{\dagger} (c_{1 \downarrow} c_{1 \uparrow})}$ 
which creates (annihilates) a double occupied state at the boundary site.
If we assume ${M_{\rm charge} \in SU(2)}$ (cf. (\ref{eq.MchargeSU2})), 
we have 
\begin{equation}
\beta = \alpha^{*},
\end{equation}
and the boundary term (\ref{eq.boundaryterma}) becomes hermitian 
(${p_{-}}$ is assumed to be real).

Similarly, we have
 \begin{eqnarray}
K_{-}^{(b)}(\theta;p_{-},M_{\rm spin})
& = & M K^{(b)}_{-}(\theta;p_{-}) M^{-1} \nonumber \\
& = & \left(
      \begin{array}{cccc}
      K_{11}(\theta) &  0      &  0     & 0 \\
      0      &  K_{22}(\theta) & K_{23}(\theta) & 0 \\
      0      &  K_{32}(\theta) & K_{33}(\theta) & 0 \\
      0      &  0      &   0    & K_{44}(\theta) 
      \end{array}
      \right), \label{eq.generalKminusb}
\end{eqnarray} 
where
\begin{eqnarray}
K_{11}(\theta) & = & K_{44}(\theta) = e^{2h} \left( 1 - 
p_{-}^2 e^{-4h} \tan^2 \theta \right),  \nonumber \\
K_{22}(\theta) & = & 1 - 2 p_{-} (M_{22} M_{33} + M_{23} M_{32}) 
\cosh 2h \tan \theta + p_{-}^2 \tan^2 \theta , \nonumber \\
K_{33}(\theta) & = & 1 + 2 p_{-} (M_{22} M_{33} + M_{23} M_{32}) 
\cosh 2h \tan \theta + p_{-}^2 \tan^2 \theta , \nonumber \\
K_{23}(\theta) & = & 4 p_{-} M_{22} M_{23} \cosh 2h \tan \theta, 
\nonumber \\
K_{32}(\theta) & = & - 4 p_{-} M_{32} M_{33} \cosh 2h \tan \theta. 
\label{eq.generalsolb}
\end{eqnarray} 
Now ${K_{-}^{(b)}(\theta;p_{-},M_{\rm spin})}$ does not depend 
on the submatrix ${M_{\rm charge}}$.
Because of the relation
\begin{equation}
M_{22} M_{33} + M_{23} M_{32} 
=  \left( 1 + 4 M_{22} M_{23} M_{32}M_{33} \right)^{\frac{1}{2}},
\end{equation}
the solution (\ref{eq.generalsolb}) depends on three arbitrary
parameters, 
\begin{equation}
p_{-}, \ \ \ \ \alpha = M_{22} M_{23}, \ \ \ \ 
\beta = - M_{32} M_{33}. \nonumber 
\end{equation} 
The corresponding boundary term is
\begin{eqnarray}
& & - p_{-} \big\{ (M_{22} M_{33} + M_{23} M_{32}) 
\left( n_{1 \uparrow} - n_{1 \downarrow} \right) 
+ 2 {\rm i} M_{22} M_{23} c_{1 \uparrow}^{\dagger} c_{1 \downarrow}
+ 2 {\rm i} M_{32} M_{33} c_{1 \downarrow}^{\dagger} c_{1 \uparrow} 
\big\} \nonumber \\
& & = - p_{-} \big\{ (1 - 4 \alpha \beta)^{\frac{1}{2}} 
\left( n_{1 \uparrow} - n_{1 \downarrow} \right) 
+ 2 {\rm i} \alpha c_{1 \uparrow}^{\dagger} c_{1 \downarrow} -  
2 {\rm i} \beta c_{1 \downarrow}^{\dagger} c_{1 \uparrow} \big\}. 
\label{eq.boundarytermb}
\end{eqnarray}
This time we see the terms ${c_{1 \uparrow}^{\dagger} c_{1 \downarrow}}$ 
and ${c_{1 \downarrow}^{\dagger} c_{1 \uparrow}}$ which flip the spin
of an electron at the boundary site.
If we assume ${M_{\rm spin} \in SU(2)}$ (cf. (\ref{eq.MspinSU2})), 
then ${\beta = \alpha^{*}}$ and the boundary term
(\ref{eq.boundarytermb}) becomes hermitian.

The obtained results can be explained by considering the ${SO(4)}$ 
rotation at the boundary site ${m=1}$, 
\begin{eqnarray}
 \left(
 \begin{array}{cc}
 \tilde{c}_{1 \downarrow}^{\dagger} & {\rm i} \tilde{c}_{1 \uparrow} \\
  {\rm i} \tilde{c}_{1 \uparrow}^{\dagger} &  \tilde{c}_{1 \downarrow} 
  \end{array}
  \right)
& = & M_{\rm spin}^{-1} 
  \left(
  \begin{array}{cc}
  c_{1 \downarrow}^{\dagger} & {\rm i} c_{1 \uparrow} \\
  {\rm i} c_{1 \uparrow}^{\dagger}   &  c_{1 \downarrow} 
    \end{array}
     \right) 
      M_{\rm charge}, \label{eq.BoundarySO(4)}
\end{eqnarray}
where we assume ${M_{\rm charge}, M_{\rm spin} \in SU(2)}$. 
The transformation (\ref{eq.BoundarySO(4)}) changes the boundary 
chemical potential into
\begin{equation}
\tilde{n}_{1 \uparrow} + \tilde{n}_{1 \downarrow} - 1 
=  \left( |M_{11}|^{2} - |M_{14}|^{2} \right) 
\left( n_{1 \uparrow} + n_{1 \downarrow} - 1 \right) 
- 2 {\rm i} M_{11} M_{14} c_{1 \uparrow}^{\dagger} 
c_{1 \downarrow}^{\dagger} + 2 {\rm i} M_{11}^{*} M_{14}^{*} 
c_{1 \downarrow} c_{1 \uparrow}, \label{eq.boundarySO(4)a}
\end{equation}
while the boundary magnetic field is changed into 
\begin{equation}
\tilde{n}_{1 \uparrow} - \tilde{n}_{1 \downarrow} 
 =  \left( |M_{22}|^{2} - |M_{23}|^{2} \right) 
\left( n_{1 \uparrow} - n_{1 \downarrow} \right) 
+ 2 {\rm i} M_{22} M_{23} c_{1 \uparrow}^{\dagger} c_{1 \downarrow} 
- 2 {\rm i} M_{22}^{*} M_{23}^{*} c_{1 \downarrow}^{\dagger} 
c_{1 \uparrow}. \label{eq.boundarySO(4)b}
\end{equation}
We see that (\ref{eq.boundarySO(4)a}) and (\ref{eq.boundarySO(4)b}) 
coincide (\ref{eq.boundaryterma}) and (\ref{eq.boundarytermb}) 
respectively. 
One may check that the Coulomb interaction term is invariant 
under the transformation (\ref{eq.BoundarySO(4)}),
\begin{equation}
\left( \tilde{n}_{1 \uparrow} - \frac{1}{2} \right) 
\left( \tilde{n}_{1 \downarrow} - \frac{1}{2} \right) 
= \left( n_{1 \uparrow} - \frac{1}{2} \right) 
\left( n_{1 \downarrow} - \frac{1}{2} \right)
\end{equation}
as it should be.

Thus we have obtained two non-diagonal solutions 
of the graded RE (\ref{eq.REK1}). 
We have confirmed that under the condition (\ref{eq.condK}), 
the solutions (\ref{eq.generalsola}) and (\ref{eq.generalsolb}) 
provide the most general solutions of the graded RE (\ref{eq.REK1}).  
It is also possible to construct the non-diagonal solutions 
of the conjugated graded RE (\ref{eq.REK2}) in a similar way.

Next we clarify the relationship between the two solutions 
${K_{-}^{(a)}(\theta;p_{-},M_{\rm charge})}$ (\ref{eq.generalKminusa})
and ${K_{-}^{(b)}(\theta;p_{-},M_{\rm spin})}$ 
(\ref{eq.generalKminusb}). 
Applying the partial particle-hole transformation of 
the fermionic ${R}$-matrix to the graded RE (\ref{eq.REK1}), 
we find that for a given solution ${K_{-}(\theta;h(\theta))}$, 
\begin{equation}
K_{-}(\theta,h(\theta)) = Q K_{-}(\theta,-h(\theta)) Q^{-1}.
\end{equation}
is also a solution of the graded RE (\ref{eq.REK1}). 
Here we have explicitly written the parameter ${h(\theta)}$. 
The transformation ${h(\theta) \rightarrow - h(\theta)}$ corresponds 
to ${U \rightarrow - U}$. 
Then the solution (\ref{eq.generalKminusb}) is 
connected with the solution (\ref{eq.generalKminusa}) 
through the following formula,
\begin{eqnarray}
K_{-}^{(b)}(\theta,h(\theta);p_{-},M_{\rm spin}) 
= Q K_{-}^{(a)}(\theta,-h(\theta);p_{-},M_{\rm charge}) Q^{-1}.
\end{eqnarray}
Here the matrix elements of ${M_{\rm charge}}$ are exchanged with 
those of ${M_{\rm spin}}$. In fact one can see that the partial 
particle-hole transformation (\ref{eq.particlehole}) changes 
the boundary term (\ref{eq.boundaryterma}) into
(\ref{eq.boundarytermb}) 
with the exchange ${M_{\rm charge} \leftrightarrow M_{\rm spin}}$. 

\vspace{30pt}
\begin{flushleft}
{\large \bf \S 7. Conclusion}
\end{flushleft}
\setcounter{equation}{0}
\renewcommand{\theequation}{7.\arabic{equation}}
In this paper we have studied the integrable boundary conditions 
for the 1D Hubbard model from the point of view of the Quantum 
Inverse Scattering Method. We have treated both the twisted periodic 
boundary condition and the open boundary condition. 
The most important object in the investigation is the fermionic 
${R}$-matrix for the 1D Hubbard model found by 
Olmedilla {\it et al}. \cite{Olmedilla1}. 
It has an interesting symmetry matrix, which reflects the ${SO(4)}$ 
symmetry of the Hamiltonian. Using the symmetry matrix, we have found 
the general twisted periodic boundary condition. 
In a sense, the periodic boundary condition can be twisted by applying 
the ${SO(4)}$ rotation to the boundary operators. 
We have also found the discrete symmetry of the fermionic
${R}$-matrix, which corresponds to the partial particle-hole 
transformation for the Hamiltonian. 
Recently the ${SO(4)}$ symmetry of the transfer matrix was
investigated by G${\ddot{\rm o}}$hmann and Murakami \cite{Gohmann}. 
The symmetry of the fermionic ${R}$-matrix is directly connected 
with the ${SO(4)}$ symmetry of the transfer matrix \cite{Shiroishi5}. 
 
For the integrable open boundary conditions, 
we have formulated the graded reflection equation in terms of 
the fermionic ${R}$-matrix. 
By solving directly the functional equations, 
we have obtained the diagonal ${K}$-matrices. 
There are two types of the ${K}$-matrices, 
which correspond to a) the boundary chemical potential and 
b) the boundary magnetic field. 
Moreover we have obtained non-diagonal ${K}$-matrices using 
the covariance property of the graded reflection equation.  
We can rotate the diagonal boundary fields by means of 
the ${SO(4)}$ symmetry. The two types of solutions are related 
through a partial particle-hole transformation 
for the fermionic ${R}$-matrix.

Recently there was reported some important progress in the evaluation 
of the eigenvalues of the transfer matrices 
of the 1D Hubbard model \cite{Ramos,Yue}. 
It is an interesting problem to generalize these results to the model 
with the twisted periodic boundary conditions 
and the open boundary condition.
   
\vspace{30pt}
\begin{flushleft}   
{\large \bf Acknowledgements}
\end{flushleft}
The authors are grateful to K. Hikami, F. G${\ddot{\rm o}}$hmann, 
S. Murakami, Y. Komori and H. Ujino for useful discussions and
comments. 
One of the authors (M. S.) also thanks X. W. Guan and A. Kuniba 
for discussions and communications. 
This work is in part supported by a Grant-in-Aid for JSPS Fellows 
from the Ministry of Education, Science, Sports and Culture of Japan.

\vspace{30pt}

\begin{flushleft}
{\large \bf Appendix: Diagonal Solutions of %
the Graded Reflection Equation }
\end{flushleft}
\setcounter{equation}{0}
\renewcommand{\theequation}{A.\arabic{equation}}
We solve the graded RE (\ref{eq.REK11}) 
for the diagonal ${K_{-}(\theta)}$
\begin{eqnarray}
& & {\cal R}_{12}(\theta_1,\theta_2) 
{\stackrel{1}{K}}_{-}(\theta_1) 
{\cal R}_{12}^{*}(\theta_1,- \theta_2) 
{\stackrel{2}{K}}_{-}(\theta_2) \nonumber \\
& & = {\stackrel{2}{K}}_{-}(\theta_2) 
{\cal R}_{12}(\theta_1, - \theta_2) 
{\stackrel{1}{K}}_{-}(\theta_1) 
{\cal R}_{12}^{*} (\theta_1,\theta_2). \label{eq.AppendixREK1}
\end{eqnarray}
Let us introduce the Boltzmann weights of 
${{\cal R}_{12}(\theta_1,-\theta_2)}$
{\scriptsize
\begin{eqnarray}
&&{\cal R}_{12}(\theta_1,- \theta_2) = \nonumber \\
&&\left( 
                               \begin{array}{cccccccccccccccc}
 \tilde{a}^{+}&0&0&0&0&0&0&0&0&0&0&0&0&0&0&0 \\
 0&-{\rm i}\tilde{b}^{+}&0&0&\tilde{e}&0&0&0&0&0&0&0&0&0&0&0 \\
 0&0&-{\rm i}\tilde{b}^{+}&0&0&0&0&0&\tilde{e}&0&0&0&0&0&0&0 \\
 0&0&0&-\tilde{c}^{+}&0&0&{\rm i}\tilde{f}&0&0&-{\rm i}\tilde{f}
&0&0&\tilde{d}^{+}&0&0&0 \\
 0&\tilde{e}&0&0&{\rm i}\tilde{b}^{-}&0&0&0&0&0&0&0&0&0&0&0 \\
 0&0&0&0&0&-\tilde{a}^{-}&0&0&0&0&0&0&0&0&0&0 \\
 0&0&0&{\rm i}\tilde{f}&0&0&\tilde{c}^{-}&0&0&-\tilde{d}^{-}
&0&0&-{\rm i}\tilde{f}&0&0&0 \\
 0&0&0&0&0&0&0&{\rm i}\tilde{b}^{-}&0&0&0&0&0&\tilde{e}&0&0 \\
 0&0&\tilde{e}&0&0&0&0&0&{\rm i}\tilde{b}^{-}&0&0&0&0&0&0&0 \\
 0&0&0&-{\rm i}\tilde{f}&0&0&-\tilde{d}^{-}&0&0&\tilde{c}^{-}
&0&0&{\rm i}\tilde{f}&0&0&0 \\
 0&0&0&0&0&0&0&0&0&0&-\tilde{a}^{-}&0&0&0&0&0 \\
 0&0&0&0&0&0&0&0&0&0&0&{\rm i}\tilde{b}^{-}&0&0&\tilde{e} &0 \\
 0&0&0&\tilde{d}^{+}&0&0&-{\rm i}\tilde{f}&0&0&{\rm i}\tilde{f}
&0&0&-\tilde{c}^{+}&0&0&0\\
 0&0&0&0&0&0&0&\tilde{e}&0&0&0&0&0&-{\rm i}\tilde{b}^{+}&0&0 \\
 0&0&0&0&0&0&0&0&0&0&0&\tilde{e}&0&0&-{\rm i}\tilde{b}^{+}&0 \\
 0&0&0&0&0&0&0&0&0&0&0&0&0&0&0&\tilde{a}^{+} \\
                      \end{array}
                      \right), \nonumber \\ \label{eq.Rmatrix2}
\end{eqnarray}
}
where
\begin{eqnarray}
\tilde{a}^{\pm} 
& = & \cos^2 (\theta_1 + \theta_2) \left\{ 1 \pm \tanh (h_1 + h_2) 
\frac{ \cos (\theta_1 - \theta_2)}{\cos (\theta_1 + \theta_2)}
\right\}, \nonumber \\
\tilde{b}^{\pm} 
& = & \sin (\theta_1 + \theta_2) \cos (\theta_1 + \theta_2) 
\left\{  1 \pm \tanh (h_1 + h_2) \frac{\sin (\theta_1 - \theta_2)}
{\sin (\theta_1 + \theta_2)} \right\} \nonumber \\
& = & \sin (\theta_1 + \theta_2) \cos (\theta_1 + \theta_2) 
\left\{ 1 \pm \tanh (h_1 - h_2) \frac{ \cos (\theta_1 - \theta_2)}
{\cos (\theta_1 + \theta_2)} \right\}, \nonumber \\
\tilde{c}^{\pm} 
& = & \sin^2 (\theta_1 + \theta_2) \left\{ 1 \pm \tanh (h_1 - h_2) 
\frac{\sin (\theta_1 - \theta_2)}{\sin (\theta_1 + \theta_2)} 
\right\}, \nonumber \\
\tilde{d}^{\pm} 
& = & 1 \pm \tanh (h_1 + h_2) \frac{\cos (\theta_1 + \theta_2)}
{\cos (\theta_1 - \theta_2)}, \nonumber \\
& = & 1 \pm \tanh(h_1 - h_2) \frac{\sin (\theta_1 + \theta_2)}
{\sin (\theta_1 - \theta_2)},  \nonumber \\      
\tilde{e}       
& = & \frac{\cos (\theta_1 + \theta_2)}{\cosh (h_1 + h_2)}, \ \ \ \ 
\tilde{f} =  \frac{\sin (\theta_1 + \theta_2)}{\cosh (h_1 - h_2)}.
\end{eqnarray}
We note useful relations among the Boltzmann weights of ${R}$-matrix,
\begin{eqnarray}
\frac{a^{\pm}}{e} 
& = & {\rm e}^{\pm (h_1 - h_2)} \cos \theta_1 \cos \theta_2 
+ {\rm e}^{\mp (h_1 - h_2)} \sin \theta_1 \sin \theta_2, \nonumber \\
\frac{c^{\pm}}{f} 
& = & {\rm e}^{\pm (h_1 + h_2)} \sin \theta_1 \cos \theta_2 
- {\rm e}^{\mp (h_1 +h_2)} \cos \theta_1 \sin \theta_2, \nonumber \\
\frac{\tilde{a}^{\pm}}{\tilde{e}} 
& = & {\rm e}^{\pm (h_1 + h_2)} \cos \theta_1 \cos \theta_2 
- {\rm e}^{\mp (h_1 + h_2)} \sin \theta_1 \sin \theta_2, \nonumber \\
\frac{\tilde{c}^{\pm}}{\tilde{f}} 
& = & {\rm e}^{\pm (h_1 - h_2)} \sin \theta_1 \cos \theta_2 
+ {\rm e}^{\mp (h_1 - h_2)} \cos \theta_1 \sin \theta_2, \nonumber \\
\frac{b^{\pm}}{e} 
& = & {\rm e}^{\pm (h_1 - h_2)} \sin \theta_1 \cos \theta_2 
- {\rm e}^{\mp (h_1 - h_2)} \cos \theta_1 \sin \theta_2 
=  \sin (\theta_1 - \theta_2) \sin (\theta_1 + \theta_2) 
\frac{\tilde{d}^{\pm}}{\tilde{f}}, \nonumber \\
\frac{b^{\pm}}{f} 
& = & {\rm e}^{\pm (h_1 + h_2)} \cos \theta_1 \cos \theta_2 
+ {\rm e}^{\mp (h_1 + h_2)} \sin \theta_1 \sin \theta_2 
= \cos (\theta_1 - \theta_2) \cos (\theta_1 + \theta_2) 
\frac{\tilde{d}^{\pm}}{\tilde{e}}, \nonumber \\
\frac{\tilde{b}^{\pm}}{\tilde{e}} & = & {\rm e}^{\pm (h_1 + h_2)} 
\sin \theta_1 \cos \theta_2 + {\rm e}^{\mp (h_1 + h_2)} 
\cos \theta_1 \sin \theta_2 
=  \sin (\theta_1 - \theta_2) \sin (\theta_1 + \theta_2) 
\frac{d^{\pm}}{f}, \nonumber \\
\frac{\tilde{b}^{\pm}}{\tilde{f}} & = & {\rm e}^{\pm (h_1 - h_2)} 
\cos \theta_1 \cos \theta_2 - {\rm e}^{\mp (h_1 - h_2)} \sin \theta_1 
\sin \theta_2 
= \cos (\theta_1 - \theta_2) \cos (\theta_1 + \theta_2) 
\frac{d^{\pm}}{e}. \nonumber \\ \label{eq.usefulrel}
\end{eqnarray}

The graded RE (\ref{eq.AppendixREK1}) for the diagonal 
${K_{-}(\theta)}$  is equivalent to the following set of 10 equations:
\begin{eqnarray}
& & x_{1}(\theta_2) \left( \frac{b^{+}}{e} x_{1}(\theta_1) 
+ \frac{\tilde{b}^{-}}{\tilde{e}} x_{2}(\theta_1) \right) 
= x_{2}(\theta_2) \left( \frac{\tilde{b}^{+}}{\tilde{e}} 
x_{1}(\theta_1) + \frac{b^{-}}{e} x_{2}(\theta_1) \right), 
\label{eq.b1} \\
& & x_{1}(\theta_2) \left( \frac{b^{+}}{e} x_{1}(\theta_1) 
+ \frac{\tilde{b}^{-}}{\tilde{e}} x_{3}(\theta_1) \right) 
= x_{3}(\theta_2) \left( \frac{\tilde{b}^{+}}{\tilde{e}} 
x_{1}(\theta_1) + \frac{b^{-}}{e} x_{3}(\theta_1) \right), 
\label{eq.b2} \\
& & x_{2}(\theta_2) \left( \frac{b^{-}}{e} x_{2}(\theta_1) 
+ \frac{\tilde{b}^{+}}{\tilde{e}} x_{4}(\theta_1) \right) 
= x_{4}(\theta_2) \left( \frac{\tilde{b}^{-}}{\tilde{e}} 
x_{2}(\theta_1) + \frac{b^{+}}{e} x_{4}(\theta_1) \right), 
\label{eq.b3} \\
& & x_{3}(\theta_2) \left( \frac{b^{-}}{e} x_{3}(\theta_1) 
+ \frac{\tilde{b}^{+}}{\tilde{e}} x_{4}(\theta_1) \right) 
= x_{4}(\theta_2) \left( \frac{\tilde{b}^{-}}{\tilde{e}} 
x_{3}(\theta_1) + \frac{b^{+}}{e} x_{4}(\theta_1) \right), 
 \label{eq.b4} \\
& & x_{1}(\theta_2) \left( x_{2}(\theta_1) + x_{3}(\theta_1) 
+ \frac{c^{+}}{f} \frac{\tilde{d}^{+}}{\tilde{f}} x_{1}(\theta_1) 
+ \frac{d^{+}}{f} \frac{\tilde{c}^{+}}{\tilde{f}} 
x_{4}(\theta_1) \right) \nonumber \\
& & = x_{4}(\theta_2) \left( x_{2}(\theta_1) + x_{3}(\theta_1) 
+ \frac{d^{+}}{f} \frac{\tilde{c}^{+}}{\tilde{f}} x_{1}(\theta_1) 
+ \frac{c^{+}}{f} \frac{\tilde{d}^{+}}{\tilde{f}} x_{4}(\theta_1)
\right), 
\label{eq.b5} \\
& & x_{2}(\theta_2) \left( x_{1}(\theta_1) + x_{4}(\theta_1) + 
\frac{c^{-}}{f} \frac{\tilde{d}^{-}}{\tilde{f}} x_{2}(\theta_1) + 
\frac{d^{-}}{f} \frac{\tilde{c}^{-}}{\tilde{f}} x_{3}(\theta_1)
\right) \nonumber \\
& & = x_{3}(\theta_2) \left( x_{1}(\theta_1) + x_{4}(\theta_1) + 
\frac{d^{-}}{f} \frac{\tilde{c}^{-}}{\tilde{f}} x_{2}(\theta_1) + 
\frac{c^{-}}{f} \frac{\tilde{d}^{-}}{\tilde{f}} x_{3}(\theta_1) 
\right), 
\label{eq.b6} \\
& & x_{1}(\theta_2) \left( \frac{\tilde{d}^{+}}{\tilde{f}} 
x_{1}(\theta_1) + \frac{{d}^{-}}{f} x_{2}(\theta_1) 
+ \frac{c^{-}}{f} x_{3}(\theta_1) + \frac{\tilde{c}^{+}}{\tilde{f}} 
x_{4}(\theta_1) \right) \nonumber \\
& & = x_{2}(\theta_2) \left( \frac{d^{+}}{f} x_{1}(\theta_1) + 
\frac{\tilde{d}^{-}}{\tilde{f}} x_{2}(\theta_1) + 
\frac{\tilde{c}^{-}}{\tilde{f}} x_{3}(\theta_1) + 
\frac{c^{+}}{f} x_{4}(\theta_1) \right), 
\label{eq.b7} \\
& & x_{1}(\theta_2) \left( \frac{\tilde{d}^{+}}{\tilde{f}} 
x_{1}(\theta_1) + \frac{{c}^{-}}{f} x_{2}(\theta_1) + 
\frac{d^{-}}{f} x_{3}(\theta) + \frac{\tilde{c}^{+}}{\tilde{f}} 
x_{4}(\theta_1) \right)  \nonumber \\
& & = x_{3}(\theta_2) \left( \frac{d^{+}}{f} x_{1}(\theta_1) + 
\frac{\tilde{c}^{-}}{\tilde{f}} x_{2}(\theta_1) + 
\frac{\tilde{d}^{-}}{\tilde{f}} x_{3}(\theta_1) + 
\frac{c^{+}}{f} x_{4}(\theta_1) \right), 
\label{eq.b8} \\
& & x_{2}(\theta_2) \left( \frac{{c}^{+}}{f} x_{1}(\theta_1) + 
\frac{\tilde{d}^{-}}{\tilde{f}} x_{2}(\theta_1) + 
\frac{\tilde{c}^{-}}{\tilde{f}} x_{3}(\theta_1) + 
\frac{d^{+}}{f} x_{4}(\theta_1) \right) \nonumber \\
& & = x_{4}(\theta_2) \left( \frac{\tilde{c}^{+}}{\tilde{f}} 
x_{1}(\theta_1) + \frac{d^{-}}{f} x_{2}(\theta_1) 
+ \frac{c^{-}}{f} x_{3}(\theta_1) + \frac{\tilde{d}^{+}}{\tilde{f}} 
x_{4}(\theta_1) \right), 
\label{eq.b9} \\
& & x_{3}(\theta_2) \left( \frac{c^{+}}{f} x_{1}(\theta_1) + 
\frac{\tilde{c}^{-}}{\tilde{f}} x_{2}(\theta_1) + 
\frac{\tilde{d}^{-}}{\tilde{f}} x_{3}(\theta_1) + 
\frac{d^{+}}{f} x_{4}(\theta_1) \right) \nonumber \\
& & = x_{4}(\theta_2) \left( \frac{\tilde{c}^{+}}{\tilde{f}} 
x_{1}(\theta_1) + \frac{{c}^{-}}{f} x_{2}(\theta_1) + 
\frac{d^{-}}{f} x_{3}(\theta_1) + \frac{\tilde{d}^{+}}{\tilde{f}} 
x_{4}(\theta_1) \right). \label{eq.b10} 
\end{eqnarray}

Due to  the relations (\ref{eq.usefulrel}), the functional equations 
(\ref{eq.b7})--(\ref{eq.b10}) reduce to 
\begin{eqnarray}
& & x_{1}(\theta_2) \left( \frac{c^{-}}{f} x_{3}(\theta_1) + 
\frac{\tilde{c}^{+}}{\tilde{f}} x_{4}(\theta_1) \right) 
= x_{2}(\theta_2) \left( \frac{\tilde{c}^{-}}{\tilde{f}}
x_{3}(\theta_1) 
+ \frac{c^{+}}{f} x_{4}(\theta_1) \right), 
\label{eq.b11} \\
& & x_{1}(\theta_2) \left( \frac{{c}^{-}}{f} x_{2}(\theta_1) 
+ \frac{\tilde{c}^{+}}{\tilde{f}} x_{4}(\theta_1) \right) = 
x_{3}(\theta_2) \left( \frac{\tilde{c}^{-}}{\tilde{f}} x_{2}(\theta_1) 
+ \frac{c^{+}}{f} x_{4}(\theta_1) \right), 
\label{eq.b12} \\
& & x_{2}(\theta_2) \left( \frac{{c}^{+}}{f} x_{1}(\theta_1) + 
\frac{\tilde{c}^{-}}{\tilde{f}} x_{3}(\theta_1) \right) = 
x_{4}(\theta_2) \left( \frac{\tilde{c}^{+}}{\tilde{f}} x_{1}(\theta_1) 
+ \frac{c^{-}}{f} x_{3}(\theta_1) \right), 
\label{eq.b13} \\
& & x_{3}(\theta_2) \left( \frac{c^{+}}{f} x_{1}(\theta_1) + 
\frac{\tilde{c}^{-}}{\tilde{f}} x_{2}(\theta_1)  \right) 
= x_{4}(\theta_2) \left( \frac{\tilde{c}^{+}}{\tilde{f}}
x_{1}(\theta_1) + \frac{{c}^{-}}{f} x_{2}(\theta_1)  \right). 
\label{eq.b14} 
\end{eqnarray}
We shall solve the functional equations (\ref{eq.b1})--(\ref{eq.b6}) 
and (\ref{eq.b11})--(\ref{eq.b14}). First, let us consider
(\ref{eq.b1}).  
Using (\ref{eq.usefulrel}), we find that (\ref{eq.b1}) is equivalent to
\begin{equation}
 \cot \theta_{1} \frac{{\rm e}^{h_{1}} 
\frac{\displaystyle x_{2}(\theta_1)}{\displaystyle x_{1}(\theta_1)} 
- {\rm e}^{- h_{1}}}{{\rm e}^{h_{1}} + {\rm e}^{-h_{1}} 
\frac{\displaystyle x_{2}(\theta_1)}{\displaystyle x_{1}(\theta_1)}} 
= \cot \theta_{2} \frac{{\rm e}^{h_{2}} 
\frac{\displaystyle x_{2}(\theta_2)}{\displaystyle x_{1}(\theta_2)} - 
{\rm e}^{- h_{2}}}{{\rm e}^{h_{2}} + 
{\rm e}^{-h_{2}} \frac{\displaystyle x_{2}(\theta_2)}
{\displaystyle x_{1}(\theta_2)}} 
 = {\rm constant}. 
\label{eq.A4}
\end{equation}
From (\ref{eq.A4}), we get
\begin{equation}
\frac{x_{2}(\theta)}{x_{1}(\theta)} 
= \frac{{\rm e}^{-h} + \alpha_{1} {\rm e}^{h} 
\tan \theta}{{\rm e}^{h} - \alpha_{1} {\rm e}^{-h} \tan \theta}, 
\label{eq.frac1}
\end{equation}
where ${\alpha_{1}}$ is a constant. 
Similarly, from (\ref{eq.b2})--(\ref{eq.b4}), we get
\begin{eqnarray}
\frac{x_{3}(\theta)}{x_{1}(\theta)} 
& = & \frac{{\rm e}^{-h} + \alpha_{2} {\rm e}^{h} \tan \theta}
{{\rm e}^{h} - \alpha_{2} {\rm e}^{-h} \tan \theta}, 
\label{eq.frac2} \\
\frac{x_{2}(\theta)}{x_{4}(\theta)} 
& = & \frac{{\rm e}^{-h} + \alpha_{3} {\rm e}^{h} \tan \theta}
{{\rm e}^{h} - \alpha_{3} {\rm e}^{-h} \tan \theta}, 
\label{eq.frac3} \\
\frac{x_{3}(\theta)}{x_{4}(\theta)} 
& = & \frac{{\rm e}^{-h} + \alpha_{4} {\rm e}^{h} \tan \theta}
{{\rm e}^{h} - \alpha_{4} {\rm e}^{-h} \tan \theta}, 
\label{eq.frac4}
\end{eqnarray}
where ${\alpha_{i} \ (i = 2,3,4)}$ are constants. 
Now from (\ref{eq.b11}) we have
\begin{eqnarray}
 & & \cot \theta_{2} \frac{{\rm e}^{h_{2}} 
\frac{\displaystyle x_{2}(\theta_2)}
{\displaystyle x_{1}(\theta_2)} - {\rm e}^{- h_{2}}}
{{\rm e}^{h_{2}} + {\rm e}^{-h_{2}} 
\frac{\displaystyle x_{2}(\theta_2)}
{\displaystyle x_{1}(\theta_2)}} 
= \cot \theta_{1} \frac{{\rm e}^{- h_{1}} - {\rm e}^{h_{1}} 
\frac{\displaystyle x_{3}(\theta_1)}
{\displaystyle x_{4}(\theta_1)}}{{\rm e}^{h_{1}} + 
{\rm e}^{-h_{1}} \frac{\displaystyle x_{3}(\theta_1)}
{\displaystyle x_{4}(\theta_1)}}. 
 \label{eq.A15}
\end{eqnarray}
Substituting (\ref{eq.frac1}) and (\ref{eq.frac4}) into
(\ref{eq.A15}), we find a constraint between the constants
\begin{equation}
\alpha_{1} = - \alpha_{4}. \label{eq.alpha14}
\end{equation}
Similarly, from (\ref{eq.b12}), we find
\begin{equation}
\alpha_{2} = - \alpha_{3}. \label{eq.alpha23}
\end{equation}
Equations (\ref{eq.b13}) and (\ref{eq.b14}) give the constraints 
(\ref{eq.alpha23}) and (\ref{eq.alpha14}) respectively. 
Finally let us consider a consistency condition 
\begin{equation}
\frac{x_{2}(\theta)}{x_{1}(\theta)}
\frac{x_{4}(\theta)}{x_{2}(\theta)} 
= \frac{x_{3}(\theta)}{x_{1}(\theta)} 
\frac{x_{4}(\theta)}{x_{3}(\theta)}. \label{eq.consistencys}
\end{equation}
 With the constraints (\ref{eq.alpha14}) and (\ref{eq.alpha23}), 
(\ref{eq.consistencys}) gives a relation
\begin{eqnarray}
\frac{{\rm e}^{-h} 
+ \alpha_{1} {\rm e}^{h} \tan \theta}{{\rm e}^{h} 
 - \alpha_{1} {\rm e}^{-h} \tan \theta} 
\frac{{\rm e}^{h} + \alpha_{2} 
{\rm e}^{-h} \tan \theta}{{\rm e}^{-h} - 
\alpha_{2} {\rm e}^{h} \tan \theta} 
= \frac{{\rm e}^{-h} 
+ \alpha_{2} {\rm e}^{h} \tan \theta}{{\rm e}^{h}
 - \alpha_{2} {\rm e}^{-h} \tan \theta} \frac{{\rm e}^{h} 
+ \alpha_{1} {\rm e}^{-h} \tan \theta}{{\rm e}^{-h} 
- \alpha_{1} {\rm e}^{h} \tan \theta}, 
\end{eqnarray}
from which we find 
\begin{equation} 
\alpha_{1} = \pm \alpha_{2}.
\end{equation}
Thus there are two possibilities 
for ${x_{i}(\theta) \ (i=1, \cdots, 4)}$. \\
a) ${\alpha_{1} = \alpha_{2} = - \alpha_{3} = - \alpha_{4} = \alpha}$
\begin{eqnarray}
\frac{x_{2}(\theta)}{x_{1}(\theta)} 
& = & \frac{{\rm e}^{-h} + \alpha {\rm e}^{h} \tan \theta}{{\rm e}^{h} 
- \alpha {\rm e}^{-h} \tan \theta}, 
\ \ \ \ \frac{x_{3}(\theta)}{x_{1}(\theta)} 
= \frac{{\rm e}^{-h} + \alpha {\rm e}^{h} \tan \theta}{{\rm e}^{h} 
- \alpha {\rm e}^{-h} \tan \theta}, \nonumber \\
\frac{x_{2}(\theta)}{x_{4}(\theta)} 
& = & \frac{{\rm e}^{-h} - \alpha {\rm e}^{h} \tan \theta}{{\rm e}^{h} 
+ \alpha {\rm e}^{-h} \tan \theta}, 
\ \ \ \ \frac{x_{3}(\theta)}{x_{4}(\theta)} = \frac{{\rm e}^{-h} - 
\alpha {\rm e}^{h} \tan \theta}{{\rm e}^{h} + \alpha {\rm e}^{-h} 
\tan \theta}. 
\label{eq.prediagonalKa}
\end{eqnarray}
b) ${\alpha_{1} = - \alpha_{2} = \alpha_{3} = - \alpha_{4} = \alpha}$
\begin{eqnarray}
\frac{x_{2}(\theta)}{x_{1}(\theta)} 
& = & \frac{{\rm e}^{-h} + \alpha {\rm e}^{h} \tan \theta}
{{\rm e}^{h} - \alpha {\rm e}^{-h} \tan \theta}, 
\ \ \ \ \frac{x_{3}(\theta)}{x_{1}(\theta)} 
= \frac{{\rm e}^{-h} - \alpha {\rm e}^{h} \tan \theta}{{\rm e}^{h} 
+ \alpha {\rm e}^{-h} \tan \theta} \nonumber \\
\frac{x_{2}(\theta)}{x_{4}(\theta)} 
& = & \frac{{\rm e}^{-h} + \alpha {\rm e}^{h} \tan \theta}{{\rm e}^{h} 
- \alpha {\rm e}^{-h} \tan \theta}, 
\ \ \ \ \frac{x_{3}(\theta)}{x_{4}(\theta)} = \frac{{\rm e}^{-h} 
- \alpha {\rm e}^{h} \tan \theta}{{\rm e}^{h} + \alpha {\rm e}^{-h} 
\tan \theta}. 
\label{eq.prediagonalKb}
\end{eqnarray}
Here ${\alpha}$ is an arbitrary constant. 
It is easy to confirm that both cases (\ref{eq.prediagonalKa}) 
and (\ref{eq.prediagonalKb}) satisfy the remaining functional 
equations (\ref{eq.b5}) and (\ref{eq.b6}). 
To conclude, we have shown that there are two diagonal solutions 
of the RE (\ref{eq.AppendixREK1}).
The solutions (\ref{eq.prediagonalKa}) and (\ref{eq.prediagonalKb}) 
are equivalent to (\ref{eq.Kminusa}) and (\ref{eq.Kminusb}), respectively.

\end{document}